# Two-dimensional Rashba semiconductors and inversion-asymmetric topological insulators in monolayer Janus $MAA'Z_xZ'_{(4-x)}$ family

Jinghui Wei[1], Qikun Tian[2], XinTing Xu[1], Guangzhao Qin[2], Xu Zuo[3], Zhenzhen Qin[1*]

[1]*Key Laboratory of Materials Physics, Ministry of Education, School of Physics, Zhengzhou University, Zhengzhou 450001, P. R. China*

[2]*State Key Laboratory of Advanced Design and Manufacturing Technology for Vehicle, College of Mechanical and Vehicle Engineering, Hunan University, Changsha 410082, P. R. China*

[3]*College of Electronic Information and Optical Engineering, Nankai University, Tianjin, 300350, P. R. China*

## Abstract

The Rashba effect in Janus structures, accompanied by nontrivial topology, plays an important role in spintronics and even photovoltaic applications. Herein, through first-principles calculations, we systematically investigate the geometric stability and electronic structures of 135 kinds of Janus $MAA'Z_xZ'_{(4-x)}$ family derived from two-dimensional $MA_2Z_4$ (M=Mg, Ga, Sr; A=Al, Ga; Z=S, Se, Te) monolayers, and design numerous Rashba semiconductors and inversion-asymmetric topological insulators. Specifically, there are a total of 26 Rashba semiconductors with isolated spin splitting bands contributed by Se/Te-$p_z$ orbitals at conduction band minimum, and the magnitude of the Rashba constant correlates strongly with both the intrinsic electric field and the strength of spin-orbit coupling (SOC). As the atomic number increases, the bandgap of Janus $MAA'Z_xZ'_{(4-x)}$ continually decreases until it shrinks to a point where, when SOC is considered, band inversion occurs, leading to a reopening of the bandgap with nontrivial topological phases. In conjunction with band inversion, $p_z$ orbitals near the Fermi level can introduce double Rashba splitting featuring a distinctive hybrid spin texture, which can be further effectively adjusted through small biaxial strains and show a continuous evolution of topological to non-topological accompanied by different spin textures. This work provides significant insights into Rashba and topology physics and further presents indispensable inversion asymmetry materials for the development of nonlinear optoelectronics.

[*] Corresponding author. E-mail: qzz@zzu.edu.cn

Spintronics[1] is a research field that has received much attention in recent years, involving technologies that utilize the spin degrees of freedom of electrons for information processing and storage[2]. The performance of spintronic devices primarily depends on the spin-orbit coupling (SOC) effect, a relativistic effect that relates the spin angular momentum of a carrier to the electrostatic potential of its environment, which has been used for spin manipulation. The Rashba effect, induced by SOC in symmetry-broken materials, locks the spin of charge carriers with their momentum [3]. This effect has attracted extensive research interest in spin field effect transistors [4–6] and has potential links to the spin Hall effect [7], spin-orbit moments [8], and nontrivial topological insulators (TIs) [9,10].

The Rashba effect can be described by the Bychkov-Rashba Hamiltonian form: $H_R = \alpha_R (\boldsymbol{\sigma}\times\boldsymbol{k})\cdot\boldsymbol{z}$, where the $\alpha_R$ represents the Rashba constant, $\boldsymbol{\sigma}$ denotes the Pauli spin matrices, $\boldsymbol{k}$ stands for the momentum, and $\boldsymbol{z}$ is the electric field direction [11]. Currently, the crucial issue to be addressed regarding the Rashba effect is to design feasible two-dimensional (2D) ideal Rashba systems and clarify the underlying mechanism of large Rashba constant in more materials [12–16]. Since the experimental synthesis of the $MoSi_2N_4$ monolayer—a member of the 2D $MA_2Z_4$ family—these materials have been theoretically predicted to exhibit significant potential in diverse fields [17–20], such as optoelectronics[21] and thermoelectricity, due to their wide bandgap, low thermal conductivity, *etc* [22–27]. Generally, constructing symmetry-broken Janus structures is an effective method for identifying Rashba semiconductors due to the introduction of intrinsic electric fields, with the crucial aspect being the presence of isolated Rashba spin-splitting [28–30]. $MA_2Z_4$ monolayers provide a natural platform for Rashba semiconductors via Janus modifications, with element-dependent spin splitting enabling tailored material selection. For the $MA_2Z_4$ family (M=Mg, Ca, Sr; A=Al, Ga; Z=S, Se, Te), $SrGa_2Se_4$ and $SrGa_2Te_4$ have been theoretically predicted as potential TIs [18]. Therefore, the Janus structures derived from the $MA_2Z_4$ family design exhibit obvious advantages in both identifying Rashba systems and exploring inversion-asymmetric TIs. It's worth mentioning that, TIs with broken inversion symmetry would substantially advance the exploration of nontrivial phenomena such as crystalline-surface-dependent topological electronic states, topological magneto-electric effects and their application as nonlinear optoelectronic spin-orbit devices [31–33], which could address a significant shortcoming of inversion-symmetric TIs [34,35]. Furthermore, Janus materials show significant potential for high nonlinear optical responses and the bulk photovoltaic effect [36–38]. The intercalation structure of the $MoS_2$-like $MZ_2$ within the $MA_2Z_4$ family offers intrinsic advantages in terms of preparative simplicity and material diversity [17,18]. Consequently, urgent exploration of Janus structures and even possible Rashba systems and inversion-symmetric TIs in 2D $MA_2Z_4$ family is needed.

In this work, we designed a series of potential Janus $MAA'Z_xZ'_{(4-x)}$ (M=Mg, Ca, Sr; A, A'=Al, Ga, In; Z $\neq$ Z'=S, Se, Te) monolayers based on the $MA_2Z_4$ and comprehensively analyzed their structural, electronic, and topological properties using first-principles calculations (See Supplementary Material Sec. A for computational details).

The top and side views of the lattice structures of $MA_2Z_4$ systems (M=Mg, Ca, Sr; A=A1, Ga, Z=S, Se, Te) are illustrated in Figure 1 (a), which consist of a seven-atomic layer with a Z–M–Z layer sandwiched between two A–Z layers. Actually, after synthesizing $MoSi_2N_4$ monolayers via chemical vapor deposition (CVD) based on $MoN_2$ monolayers [17], subsequent studies have explored the feasibility of synthesizing other $MA_2Z_4$ monolayers in such intercalated architecture [17,18]. Taking $MgAl_2S_4$ monolayer as an example, by comparing the total energy of different phases (Table S5), we find that for $MA_2Z_4$ monolayer with M as Mg, Ca or Sr, the hexagonal lattice structure with $P\bar{3}m1$ space group formed by M, A and Z atoms is the most energetically favorable in $\beta_2$ phase. Our results are consistent with previous studies [18]. Inspired by the intercalated architecture of the $MA_2Z_4$ family, possible Janus configurations could be naturally designed based on the 2D pure $\beta_2$-$MA_2Z_4$ systems, that is, $MZ_2$ or Janus-MZZ' inserted between the AZ layer and A'Z' layer, as shown in Figure 1 (b). Notably, the unique 'sandwich structure' of $MA_2Z_4$ monolayers enables easy manufacturing of these Janus configurations in practice. As shown in Figure 1 (c), we design seven kinds of monolayer Janus $MAA'Z_xZ'_{(4-x)}$, and these configurations, with M as the central atom, can be divided into '7/5/3-Janus' configurations. After considering all Janus configurations and elemental substitutions, we identified a total of 135 distinct $MAA'Z_xZ'_{(4-x)}$ monolayer configurations (M=Mg, Ca, Sr; A, A'=Al, Ga, In; Z, Z'=S, Se, Te) with a non-centrosymmetric $P3m1$ space group. Among these configurations, there are 36 of the '7-Janus' type, 36 of the '5-Janus' type, and 63 of the '3-Janus' type, respectively. Lattice parameters are summarized in Tables S1-S4. The lattice parameters of the $MA_2Z_4$ monolayers range from 3.68 Å ($MgAl_2S_4$) to 4.37 Å ($SrGa_2Te_4$), which is basically consistent with previous work (3.71-4.41 Å) [18]. Correspondingly, the lattice parameters of $MAA'Z_xZ'_{(4-x)}$ monolayers vary from 3.70 to 4.46 Å, increasing with the increase of the atomic radius of the doped elements.

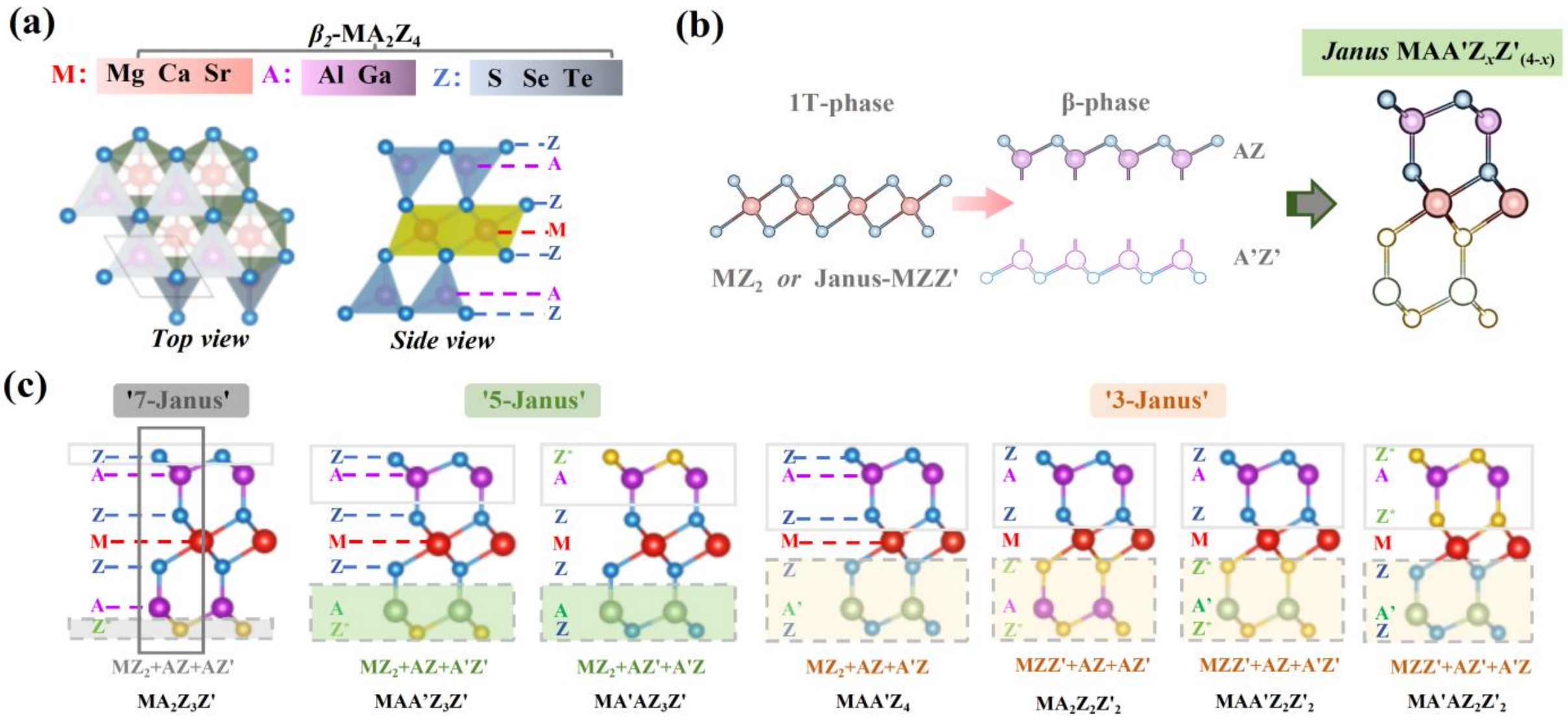

**Figure 1** (a) Top and side views of lattice structures of $MA_2Z_4$ systems. (b) 2D Janus $MAA'Z_xZ'_{(4\text{-}x)}$ with sandwich structure constructed by $MZ_2$ and $A_2Z_2$ layers. (c) Side views of lattice structures of the $MAA'Z_xZ'_{(4\text{-}x)}$ systems. The three Janus configurations based on the seven-atomic layer structure are: '7-Janus', '5-Janus', and '3-Janus'. Inside the dotted box are the replaced atomic layers.

To confirm the structural stability of $MAA'Z_xZ'_{(4\text{-}x)}$ systems, we take $MgAA'Z_xZ'_{(4\text{-}x)}$ monolayers as an example and calculate their cohesion energy $E_{\mathrm{coh}}$ by the following formula:

$$E_{\mathrm{coh}} = [E_{\mathrm{tot}} - E_{\mathrm{M}} - E_{\mathrm{A}} - E_{\mathrm{A'}} - \mathrm{x}E_{\mathrm{Z}} - (4-\mathrm{x})E_{\mathrm{Z'}}]/7$$

where $E_{\mathrm{tot}}$ is the total energy of the monolayers; $E_M$, $E_A$, $E_{A'}$, $E_Z$ and $E_{Z'}$ are the monatomic energies of M, A, A', Z and Z', respectively; x is the number of Z atoms in the unit cell. As shown in Figure 2 (a), all $E_{coh}$ values are negative, indicating the Janus $MgAA'Z_xZ'_{(4\text{-}x)}$ monolayers are energy stable. In addition, the $E_{\mathrm{coh}}$ of the pure $MA_2Z_4$ are also shown in Figure 2 (a), demonstrating its stability. Similarly, ab initio molecular dynamics (AIMD) simulations at 300 K are conducted for both the lightest and heaviest monolayers of the $MgAA'Z_xZ'_{(4\text{-}x)}$ systems in different '7/5/3-Janus' configurations to verify their thermal stability, as illustrated in Figure 2 (b). The free energy fluctuates in a narrow range, with no significant distortion of the equilibrium structure or bond breaking, indicating that the $MgAA'Z_xZ'_{(4\text{-}x)}$ monolayers are thermal stable. In addition, the phonon spectra presented in Figure 2 (c) and S1 demonstrate that the $MgAA'Z_xZ'_{(4\text{-}x)}$ monolayers possess three acoustic and eighteen optical phonon branches devoid of imaginary vibration frequencies, thereby confirming their dynamic stability.

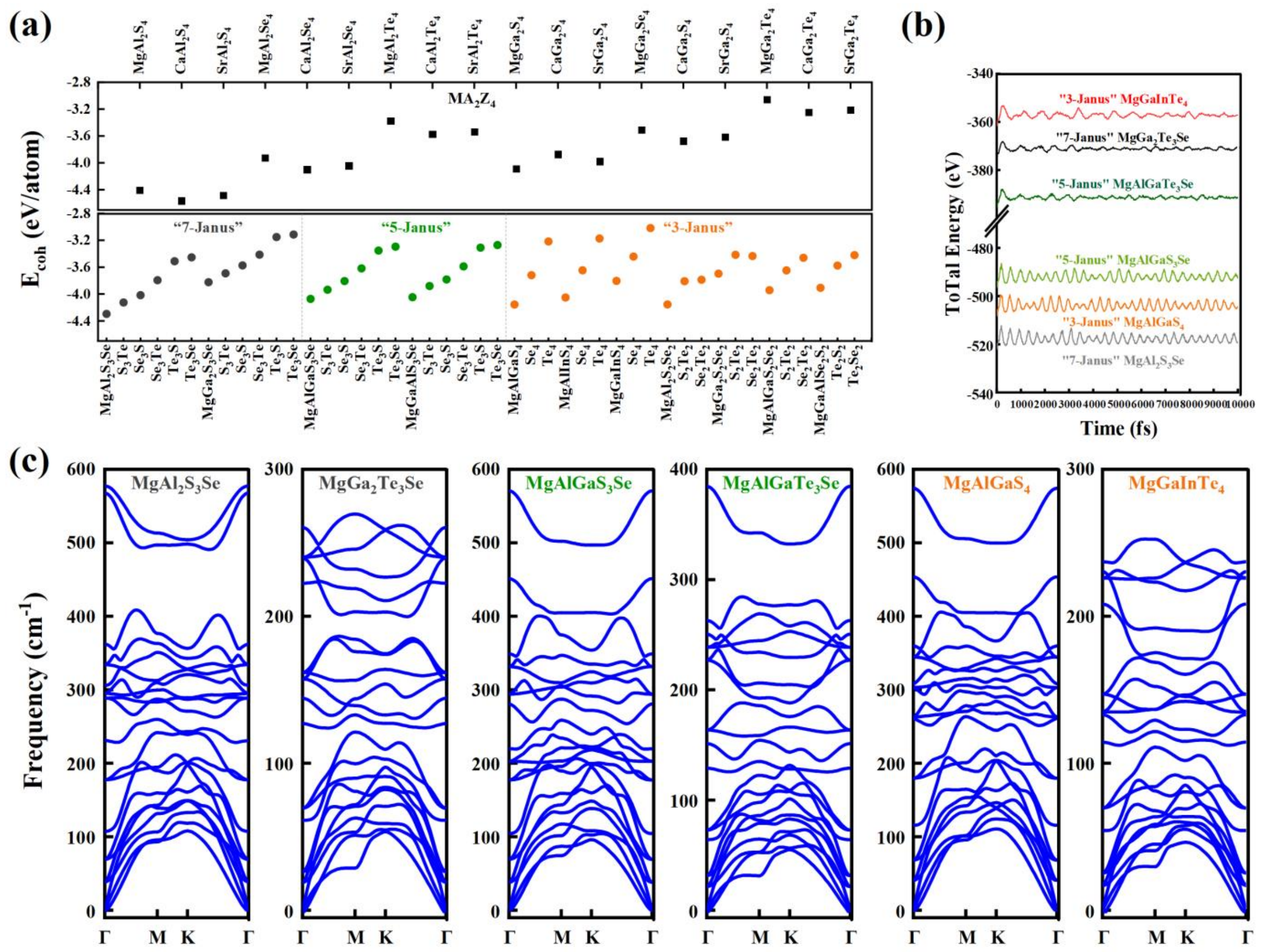

**Figure 2** (a) Cohesion energy $E_{coh}$ of $MA_2Z_4$ monolayers and MgAA'$Z_xZ'_{(4-x)}$ monolayers. (b) AIMD simulations of total energy fluctuation under 10000 fs at 300 K and (c) Phonon spectra of six MAA'$Z_xZ'_{(4-x)}$ monolayers.

The electronic properties of $MA_2Z_4$ and MAA'$Z_xZ'_{(4-x)}$ systems are investigated with and without the SOC in the Perdew-Burke-Ernzerhof (PBE) functional [39]. As shown in Figure 3 (a), when the central atom M is Mg, the bandgaps of the '3-Janus' MgAA'$Z_4$ monolayers progressively decrease from 1.57 to 0.14 eV upon substitution of Z atoms from S to Te and AA' atoms from AlGa to GaIn, indicating that the introduction of heavier elements has a significant impact on the bandgap of the material. The contribution of conduction band minimum (CBM) and valence band maximum (VBM) of these systems is mainly attributed to their A-$s$, Z-$p_z$ and Z-$p_{x,y}$ orbitals, respectively, suggesting that the atomic-orbital contribution near the Fermi level is not substantially affected by different doping elements. When the central atom M changes from Mg to Ca or Sr, the bandgap and orbital are similar to those of MgAA'$Z_4$, as shown in Figure S5. Tables S1 to S4 provide the bandgaps of all $MA_2Z_4$ and Janus MAA'$Z_xZ'_{(4-x)}$ monolayers. The bandgap range of the MAA'$Z_xZ'_{(4-x)}$ systems is also 0 to 2.17 eV, and most semiconductors have a direct bandgap (See Supplementary Material Sec. B for details).

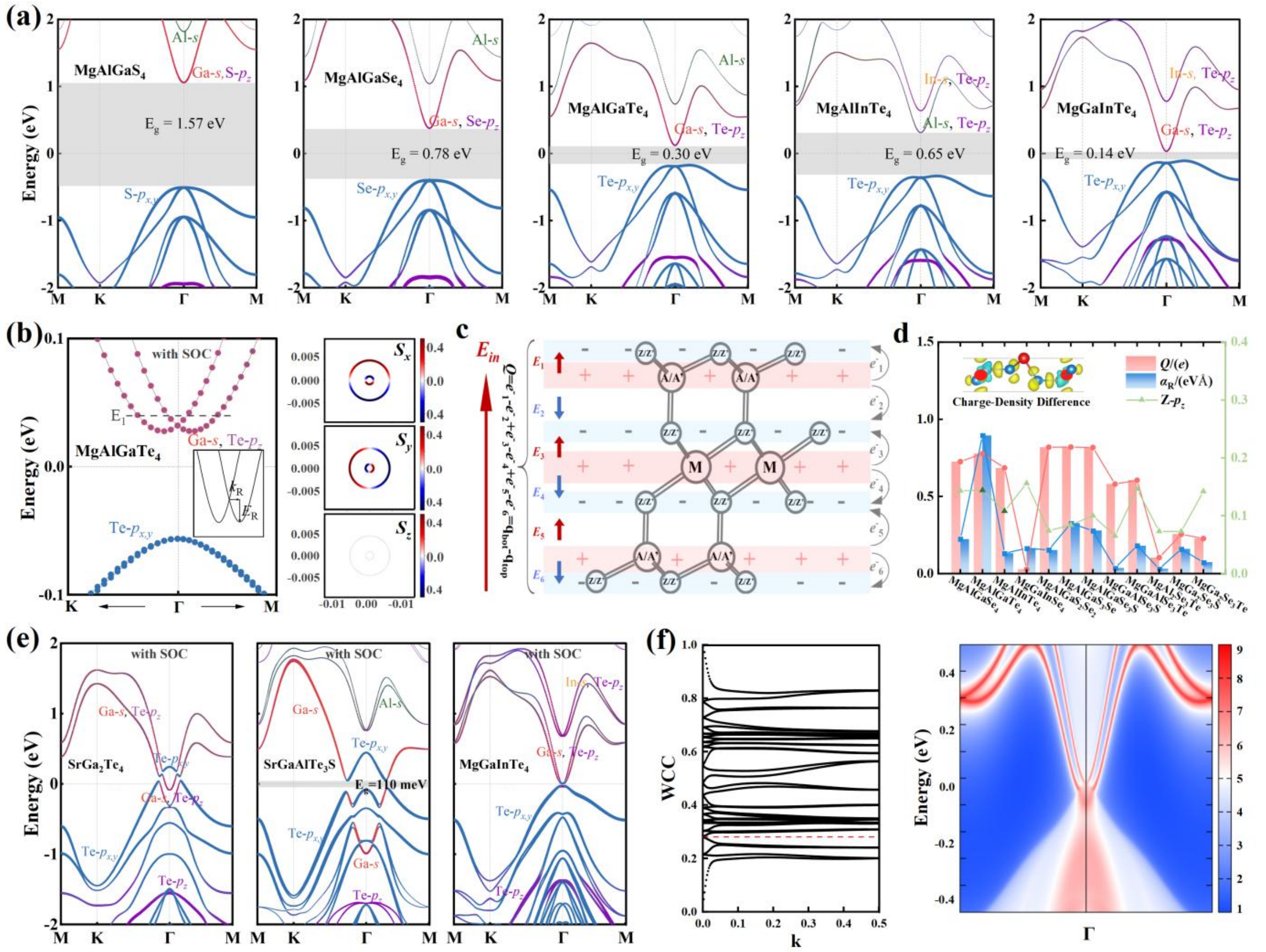

**Figure 3** (a) Orbital-projected band structures of $MgAA'Z_4$ monolayers without considering SOC. Different colors represent different atomic-orbital contribution. (b) The projected band structure near the Fermi level of $MgAlGaTe_4$ monolayer when SOC is considered. An enlarged view of the Rashba spin-splitting band at CBM is plotted within it. The right parts represent 2D contour plots of spin textures for the $MgAlGaTe_4$ monolayer at constant energy $E_1$ ($E_f$ + 0.04 eV) in $k_x - k_y$ plane centered at Γ point. Red color represents spin-up states and blue represents spin-down states. Spin texture projections along the $S_x$, $S_y$, and $S_z$ directions are plotted. (c) Schematic diagrams of the charge transfer and $E_{in}$ for the $MAAZ_xZ'_{(4\text{-}x)}$ monolayers. (d) The total charge transfer $Q$ of the $MgAA'Z_xZ'_{(4\text{-}x)}$ monolayers with Rashba effect and the contribution of the corresponding $Z$-$p_z$ orbital. (e) The projected band structures of $SrGa_2Te_4$, $SrGaAlTe_3S$, $MgGaInTe_4$ monolayers considering SOC. (f) The WCC evolution and edge state of the $MgGaInTe_4$ monolayer.

Furthermore, most of the $MAA'Z_xZ'_{(4\text{-}x)}$ semiconductors exhibit Rashba splitting at CBM when SOC is considered, despite their possession of different Janus configurations and various element compositions. To quantify the strength of the Rashba effect, we calculate the Rashba constant $\alpha_R$ for all 26 Rashba semiconductors with $\alpha_R \geq 0.025$ eV·Å by $\alpha_R = 2E_R/k_R$, where $E_R$ is the Rashba energy and $k_R$ is the momentum offset, with $MgAlGaTe_4$ having the largest Rashba constant (0.89 eV·Å). Using $MgAlGaTe_4$ monolayer as an illustrative example, we further observe the effect of Rashba spin splitting through spin texture calculation, whereby the opposite chiral circles of *in-plane* spin projection ($S_x$ and $S_y$) in Figure 3 (b) confirms our results. The projected band structure in Figure 3 (b) indicates that the VBM of $MgAlGaTe_4$ monolayer is mainly contributed by Te-$p_{x,\,y}$ orbitals, while the CBM exhibiting Rashba splitting is mainly contributed by Ga-*s* and Te-$p_z$ orbitals. In addition, the intrinsic electric field ($E_{in}$) and charge transfer ($Q$) indicate the presence of localized $E$ and $Q$ between each atomic layer within the MAA'$Z_x$Z'$_{(4-x)}$ monolayer as shown in Figure 3 (c). Quantitative analysis reveals that $\alpha_R$ scales linearly with both the $Q$ and the SOC strength, as shown in Figure 3 (d) and Table S7 (See Supplementary Material Sec. D for details). In addition, all the Rashba splitting bands have contributions from the Z-$p_z$ orbital.

With further doping of heavy elements such as Te, the bandgap of Janus MAA'$Z_x$Z'$_{(4-x)}$ decreases continuously until the bandgap, when combined with strong spin-orbit coupling, becomes small enough to lead to band inversion with a re-opened gap ($10 \leq E_{g\text{-}soc} \leq 110$ meV), transitioning from trivial Rashba semiconductors to nontrivial TIs. As shown in Figure 3 (e), when considering SOC, both the band structures of the $SrGaAlTe_3S$ and $MgGaInTe_4$ monolayers exhibit band inversion, with global bandgaps of 110 meV and 27 meV, respectively. The projected band structures indicate that the Ga-*s* and Te-*p* orbitals are inverted around Γ point near the Fermi level, where the *s*-*p* band inversion strongly points to the nontrivial topological phases. In addition, as shown in Figures S3 and S4, the band structures of the remaining Janus MAA'$Z_x$Z'$_{(4-x)}$ also show a similar band inversion around the Fermi level with varying global bandgaps, indicating the presence of inversion-asymmetric nontrivial topological phases. To verify the nontrivial topology of the above Janus MAA'$Z_x$Z'$_{(4-x)}$ with band inversion, we further plot the Wannier charge center (WCC) evolution of $MgGaInTe_4$ monolayer in Figure 3 (f), which shows that for any horizontal reference line (e.g., WCC=0.25), there are an odd number of crossing points, indicating that it is a TI with $Z2=1$. The topological edge state of Figure 3 (f) further confirms its nontrivial topology. Similarly, the nontrivial topology of other Janus monolayers, such as the $CaAlGaTe_4$ monolayer, is also confirmed by its WCC evolution in Figure S6, verifying the presence of inversion-asymmetric nontrivial topological phases. It is noteworthy that these asymmetric Janus TIs, exhibiting broad bands near the Fermi level contributed by *s* and *p* orbitals, enable asymmetric photoexcited carriers and generate more nonzero shift current polarization components, boosting their photovoltaic performance and promising bulk photovoltaic effect material discovery [37,38].

In most Janus MAA'$Z_x$Z'$_{(4-x)}$ TIs, the band inversion is often accompanied by Rashba splitting at the CBM, thus resulting in a unique hybrid spin texture at the VBM. Figure 4 (a) shows the hybrid spin texture in the $S_x$ direction of the $MgGaInTe_4$ monolayer. Unusually, both $E_2$ and $E_3$ within the valence band exhibit a consistent spin texture on the same side, while differing on opposing sides, which is different from the typical Rashba spin texture. By correlating the spin texture of the valence band with the projected band structure, it can be observed the inner and outer layers of the texture originate from bands contributed by Ga-$s$, Te-$p_z$, and Te-$p_{x,y}$, respectively. Notably, the band arising from Te-$p_{x,y}$ orbitals not only introduces a significant *out-of-plane* spin component but may also account for the unusual arrangement of spin-up and spin-down textures, as shown in Figure S7. Additionally, the nontrivial topological states of $MAA'Z_xZ'_{(4-x)}$ systems can be further effectively adjusted through small biaxial strains and show a continuous evolution of topological to non-topological accompanied by different spin textures. Taking $MgGaInTe_4$ as an example, when subjected to -2% biaxial strain, it transforms into a Rashba semiconductor. Under 2% biaxial strain, it remains a TI but with altered spin textures, as shown in Figure 4 (a) and Figure S7 (See Supplementary Material Sec. D for details of spin texture evolution). The schematic diagram in Figure 4 (b) clearly illustrates a series of changes in the bands and spin textures of the $MgGaInTe_4$ monolayer under small strains, the variations in the $s$, $p_z$, and $p_{x,y}$ orbitals near the Fermi level are the triggers behind the band inversion and hybrid textures. The evolution of atomic orbitals near the Fermi level in the right-most subfigure of Figure 4 (b) has two distinct stages. In stage (I), the $s$ and $p$ orbitals split into $|s^{\pm}>$ and $|p^{\pm}>$ due to chemical bonding and crystal field effects, where the superscripts $\pm$ indicating the bond and antibonding states. The band near the Fermi level is mainly composed of $|s^{-}>$ , $|p_z^{-}>$ and $|p_{x,y}^{+}>$. In stage (II), when the energy difference between $|s^{-}>$ , $|p_z^{-}>$ and $|p_{x,y}^{+}>$ is small enough, with the SOC effect, $|s^{-}>$ , $|p_z^{-}>$ exchanges positions with $|p_{x,y}^{+}>$ and causes band inversion. Previous research indicates that decoding spin structures [40,41] and manipulating spin textures using twisting angles [42–44] can generate information related to the electronic structure, paving the way for practical applications of spin-based devices. In this work, doping and strain can also effectively tune the spin textures in $MAA'Z_xZ'_{(4-x)}$ systems, offering alternative methods for spin texture modulation.

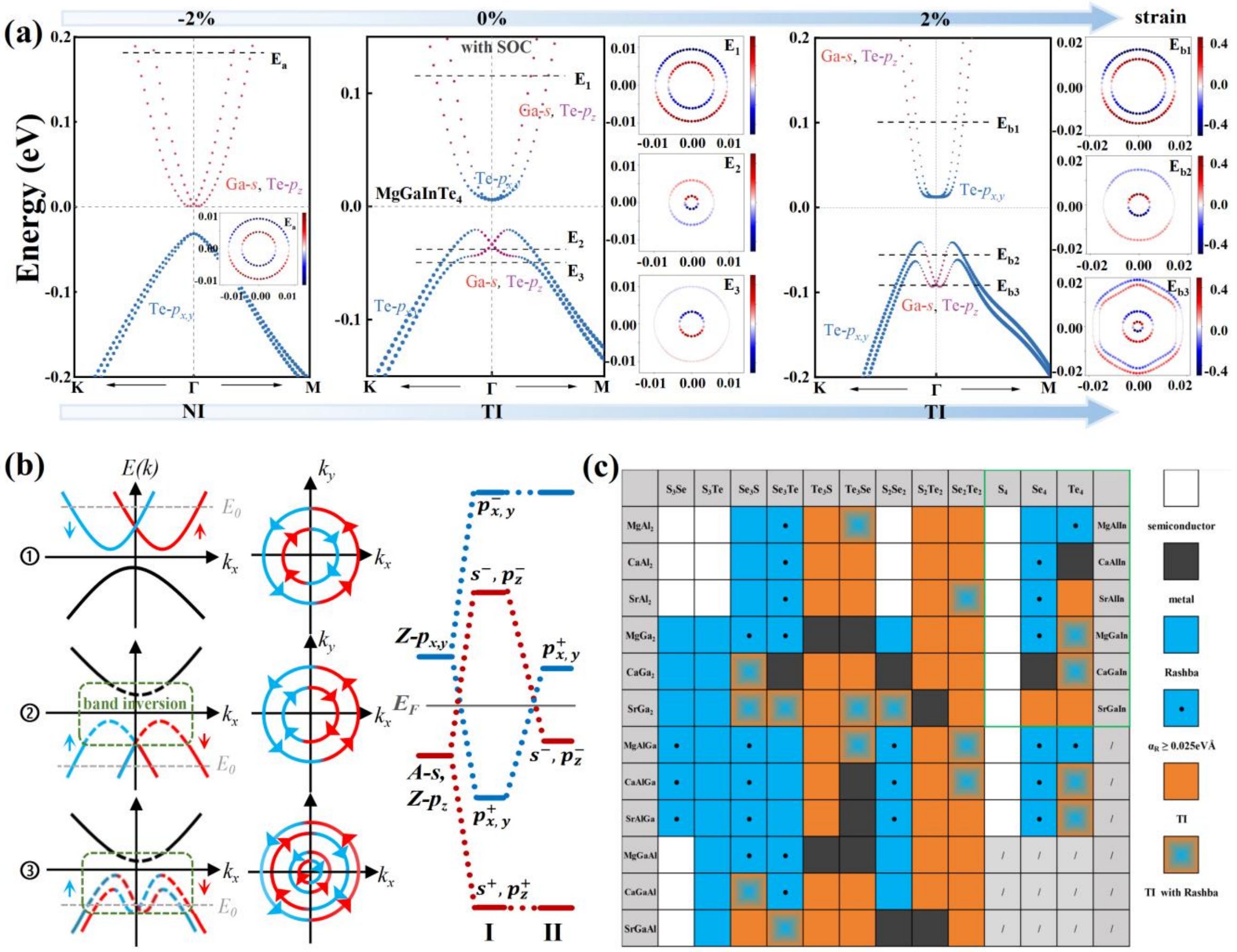

**Figure 4** (a) The orbital-projection band structures and spin textures of $MgGaInTe_4$ monolayer within 2% biaxial strain when considering SOC. (b) Three different Rashba spin splitting (left) and texture (middle) diagrams, as well as orbital variation (right) diagram. (c) Schematic diagram of the electronic structure characteristics of all monolayers in the MAA'$Z_xZ'_{(4-x)}$ systems when SOC is considered. The horizontal axis (first row) and the vertical axis (first column) represent the MAA' and $Z_xZ'_{(4-x)}$ composed of different elements, respectively.

To establish the relationship between the atomic composition and electronic properties of MAA'$Z_xZ'_{(4-x)}$, we present a table summarizing their electronic structures in Figure 4 (c), and statistically analyze the orbital contributions near the Fermi level at the Γ point in Figure S11. The results indicate that the emergence of the Rashba effect and topological states is closely related to the elemental composition (See Supplementary Material Sec. D for details). All Rashba semiconductors exhibit contributions from Se-$p_z$ or Te-$p_z$ orbitals near the Fermi level at the Γ point. Due to the intrinsic electric field along the Z direction in the Janus MAA'$Z_xZ'_{(4-x)}$ systems, $\langle p_z|E_z|p_z\rangle \neq 0$, while $\langle p_{x,\ y}|E_z|p_{x,y}\rangle = 0$, the contribution of the $p_z$ orbital of the Z atom is the primary factor responsible for the occurrence of the Rashba effect at the CBM [45,46]. Moreover, nontrivial topological states predominantly emerge in systems containing Te elements, accompanied by *s*-*p* orbital hybridization, which also reflects the design principles for inversion-asymmetric TIs. The MAA'$Z_xZ'_{(4-x)}$ systems not only enable the transition from trivial Rashba semiconductors to TIs through doping but also allow for the modulation of its nontrivial topological states and spin textures via strain. This tunable phase-transition Janus TIs discussed may have potential applications in fields such as nonlinear optoelectronics.

In summary, we designed numerous Rashba semiconductors and inversion-asymmetry nontrivial TIs via the specific '7/5/3-Janus' forms in a series of $MA_2Z_4$ monolayers, and focused on the evolution under different combinations as well as the physical mechanism. Such 135 kinds of Janus MAA'$Z_xZ'_{(4-x)}$ family are confirmed dynamically and thermally stable, indicating their potential for experimental synthesis. Specifically, 26 Rashba semiconductors exhibit isolated spin-splitting bands dominated by Se/Te-$p_z$ orbitals at the CBM. However, the $\alpha_R$ values range from 0.025 eVÅ to 0.894 eVÅ, due to variations in the intrinsic vertical electric field and the strength of SOC across different systems. Actually, as the total atomic number rises, the bandgaps of Janus MAA'$Z_xZ'_{(4-x)}$ decrease continuously and the trend continues until the bandgap becomes sufficiently small to cause band inversion, with a reopened gap ($10 \leq E_{g\text{-}soc} \leq 110$ meV) when SOC is included, exhibiting nontrivial topology in 54 systems. In conjunction with band inversion, $p_z$ orbitals near the Fermi level give rise to double Rashba splitting, accompanied by a distinct hybrid spin texture. This can be further effectively adjusted through small biaxial strains, showing a continuous evolution from topological to non-topological states, accompanied by different spin textures. Our designed Janus MAA'$Z_xZ'_{(4-x)}$ family encompasses multiple Rashba semiconductors and inversion-asymmetry TIs, showcasing the obvious evolution from topological to non-topological states in different Janus structures and when applying small strains, particularly accompanied by traceable Rashba spin textures. Our work not only expands the diverse Janus MAA'$Z_xZ'_{(4-x)}$ family with multifunctional application prospects but also reveals the designing rules of Rashba semiconductors and inversion-asymmetric topological insulators.

See the supplementary material for more details on computational methods, phonon dispersion, band structures, spin textures and analysis of intrinsic electric field.

## ACKNOWLEDGEMENTS

Z. Q. is supported by the National Natural Science Foundation of China (Grant No.12274374). G. Q. is supported by the National Natural Science Foundation of China (Grant No. 52006057), the Fundamental Research Funds for the Central Universities (Grant No.

531119200237 and 541109010001). The numerical calculations in this work are supported by National Supercomputing Center in Zhengzhou.

## AUTHOR CONTRIBUTIONS

Z.Q. conceived and designed the research. J.W. carried out the calculations and analysed the calculated results. All the authors contributed to the final revision of this article.

## COMPETING INTERESTS

The authors declare no competing interests.

## DATA AVAILABILITY

The data that supports the findings of this study are available within the article and its supplementary material.

*Supplemental Material*

# Two-dimensional Rashba semiconductors and inversion-asymmetric topological insulators in monolayer Janus MAA'$Z_x$Z'$_{(4-x)}$ family

Jinghui Wei[1], Qikun Tian[2], XinTing Xu[1], Guangzhao Qin[2], Xu Zuo[3], Zhenzhen Qin[1*]

[1]*Key Laboratory of Materials Physics, Ministry of Education, School of Physics, Zhengzhou University, Zhengzhou 450001, P. R. China*

[2]*State Key Laboratory of Advanced Design and Manufacturing Technology for Vehicle, College of Mechanical and Vehicle Engineering, Hunan University, Changsha 410082, P. R. China*

[3]*College of Electronic Information and Optical Engineering, Nankai University, Tianjin, 300350, P. R. China*

[*] Corresponding author. E-mail: qzz@zzu.edu.cn

## A. COMPUTATIONAL METHODS

First-principles calculations were performed by applying the Vienna Ab initio Simulation Package (VASP) based on density-functional theory (DFT) and the projected-augmented wave method [1,2]. The exchange-correlation functional was processed in the generalized gradient approximation (GGA) of Perdew-Burke-Ernzerhof (PBE) [3]. We relaxed all atoms to ensure an energy convergence of $10^{-6}$ eV and forcing convergence of $10^{-4}$ eV/Å. In the calculation, a 15 × 15 × 1 *k*-point grid centered on Gamma was used to sample the Brillouin zone (BZ). The thickness of the vacuum layer was set to ~ 20 Å along the *z* direction to eliminate the interaction between adjacent layers. To verify the stability of $MAA'Z_xZ'_{(4-x)}$ monolayers, we generated a 4 × 4 × 1 supercell for phonon spectrum calculation, which was performed using PHONOPY code [4]. Moreover, we tested the thermal stability of $MAA'Z_xZ'_{(4-x)}$ monolayers by performing ab initio molecular dynamics (AIMD) simulations at room temperature for 10000 fs with a step of 2 fs. The number of *k*-points per path segment was set to 200. The Bader techniques were used to analyze the charge transfer [5]. The constant-energy contour plots of the spin texture were plotted using the PYPROCAR code [6]. Using Wannier90 and WannierTools codes, a tight bound Hamiltonian with the largest local Wannier function was constructed to fit the band structure, and then the $Z_2$ invariants were calculated [7,8], which can be confirmed via calculations of the Wannier charge center (WCC).

## B. Structural parameters and stability

The Janus $MAA'Z_xZ'_{(4-x)}$ family can be categorized into '7/5/3-Janus' configurations as follows: 1. For the '7-Janus' type, only the outermost atom on one side is selected for replacement ($MZ_2$+AZ+AZ'); 2. For the '5-Janus' type, with (ZA)ZMZ(AZ) as matrix, A and Z on one side are selected for diatomic layer replacement ($MZ_2$+AZ+A'Z', $MZ_2$+AZ'+A'Z); 3. For the '3-Janus' type, take (ZAZ)M(ZAZ) as the matrix and select the ZAZ atom on one side for single or synchronous replacement of A and ZZ ($MZ_2$+AZ+A'Z, MZZ'+AZ+AZ', MZZ'+AZ+A'Z', MZZ'+AZ'+A'Z).

The lattice parameters and bandgap for $MA_2Z_4$ monolayers are provided in Table S1, while those of the '7/5/3-Janus' $MAA'Z_xZ'_{(4-x)}$ monolayers are given in Tables S2 to S4. In addition, phonon spectra of all $MgAA'Z_xZ'_{(4-x)}$ monolayers are provided. The bandgap range of the $MA_2Z_4$ systems, without considering spin-orbit coupling (SOC), is from 0 to 2.17 eV, which is generally consistent with the previous study (0 to 2.08 eV) [9]. The lattice parameter (2.90 Å) and bandgap (1.79 eV) of $MoSi_2N_4$ monolayer, as calculated in this study, are in agreement with both theoretical values (2.91 Å, 1.74 eV) and experimental results (2.94 Å, 1.94 eV), as reported in Ref. [9] and [10], respectively. The bandgaps of $MAA'Z_xZ'_{(4-x)}$ systems tend to decrease with the increase of the atomic number of the constituent elements. Among these, the $MgAl_2S_3Se$ monolayer exhibits a maximum bandgap of 2.17 eV, showcasing semiconductor behavior, while some materials have a bandgap of 0 eV, exhibiting metallic properties. Most $MAA'Z_xZ'_{(4-x)}$ semiconductors have a direct bandgap, with CBM and VBM located at the Γ point, except for these seven monolayers, $MgAl_2Se_3Te$, $MgGaAlSe_3Te$, $MgAlGaTe_4$, $MgAlInSe_4$, $MgAlInTe_4$, $MgGaInSe_4$ and $MgGaInTe_4$, which are indirect bandgap semiconductors. When SOC is further taken into account, the bandgaps of $MAA'Z_xZ'_{(4-x)}$ systems are mostly reduced even more, as shown in Tables S2 to S4.

**Table S1**. Lattice parameters and bandgaps of 2D $MA_2Z_4$ (M=Mg, Ga, Sr; A=Al, Ga; Z=S, Se, Te) systems.

| Classes | Systems | Structure | Phase | $a=b$ (Å) | $E_{g\text{-PBE}}$ (eV) | $E_{g\text{-PBE-SOC}}$ (eV) |
|---|---|---|---|---|---|---|
| $MAl_2Z_4$ | $MgAl_2S_4$ | $P\bar{3}m1$ | $\beta_2$ | 3.68 | 2.17 | 2.13 |
| | $CaAl_2S_4$ | $P\bar{3}m1$ | $\beta_2$ | 3.77 | 1.79 | 1.76 |
| | $SrAl_2S_4$ | $P\bar{3}m1$ | $\beta_2$ | 3.83 | 1.59 | 1.56 |
| | $MgAl_2Se_4$ | $P\bar{3}m1$ | $\beta_2$ | 3.89 | 1.33 | 1.20 |
| | $CaAl_2Se_4$ | $P\bar{3}m1$ | $\beta_2$ | 3.98 | 0.94 | 0.81 |
| | $SrAl_2Se_4$ | $P\bar{3}m1$ | $\beta_2$ | 4.02 | 0.79 | 0.66 |
| | $MgAl_2Te_4$ | $P\bar{3}m1$ | $\beta_2$ | 4.23 | 0.82 | 0.58 |
| | $CaAl_2Te_4$ | $P\bar{3}m1$ | $\beta_2$ | 4.30 | 0.43 | 0.18 |
| | $SrAl_2Te_4$ | $P\bar{3}m1$ | $\beta_2$ | 4.35 | 0.31 | 0.06 |
| $MGa_2Z_4$ | $MgGa_2S_4$ | $P\bar{3}m1$ | $\beta_2$ | 3.71 | 1.38 | 1.35 |
| | $CaGa_2S_4$ | $P\bar{3}m1$ | $\beta_2$ | 3.80 | 0.87 | 0.84 |
| | $SrGa_2S_4$ | $P\bar{3}m1$ | $\beta_2$ | 3.86 | 0.68 | 0.64 |
| | $MgGa_2Se_4$ | $P\bar{3}m1$ | $\beta_2$ | 3.91 | 0.60 | 0.47 |
| | $CaGa_2Se_4$ | $P\bar{3}m1$ | $\beta_2$ | 4.00 | 0.15 | 0.00 |
| | $SrGa_2Se_4$ | $P\bar{3}m1$ | $\beta_2$ | 4.05 | 0.00 | 0.05 |
| | $MgGa_2Te_4$ | $P\bar{3}m1$ | $\beta_2$ | 4.25 | 0.15 | 0.05 |
| | $CaGa_2Te_4$ | $P\bar{3}m1$ | $\beta_2$ | 4.32 | 0.00 | 0.00 |
| | $SrGa_2Te_4$ | $P\bar{3}m1$ | $\beta_2$ | 4.37 | 0.00 | 0.00 |

**Table S2**. Lattice parameters, bandgaps and Rashba parameters of "7-Janus" configuration in 2D $MAA'Z_2Z'_2$ systems.

| Classes | Systems | Structure | Phase | $a=b$ (Å) | $E_{g\text{-PBE}}$ (eV) | $E_{g\text{-PBE-SOC}}$ (eV) | $E_R$ (eV) | $k_R$ (Å$^{-1}$) | $\alpha_R$ (eVÅ) |
|---|---|---|---|---|---|---|---|---|---|
| $MAl_2Z_3Z'$ | $MgAl_2S_3Se$ | $P3m1$ | $\beta_2$ | 3.75 | 2.17 | 2.14 | \ | \ | \ |
| | $CaAl_2S_3Se$ | $P3m1$ | $\beta_2$ | 3.84 | 1.73 | 1.69 | \ | \ | \ |
| | $SrAl_2S_3Se$ | $P3m1$ | $\beta_2$ | 3.90 | 1.55 | 1.51 | \ | \ | \ |
| | $MgAl_2S_3Te$ | $P3m1$ | $\beta_2$ | 3.86 | 1.33 | 1.05 | \ | \ | \ |
| | $CaAl_2S_3Te$ | $P3m1$ | $\beta_2$ | 3.96 | 1.23 | 0.99 | \ | \ | \ |
| | $SrAl_2S_3Te$ | $P3m1$ | $\beta_2$ | 4.02 | 1.14 | 0.93 | \ | \ | \ |
| | $MgAl_2Se_3S$ | $P3m1$ | $\beta_2$ | 3.83 | 1.06 | 0.93 | \ | \ | \ |
| | $CaAl_2Se_3S$ | $P3m1$ | $\beta_2$ | 3.90 | 0.71 | 0.58 | \ | \ | \ |
| | $SrAl_2Se_3S$ | $P3m1$ | $\beta_2$ | 3.95 | 0.58 | 0.45 | \ | \ | \ |
| | $MgAl_2Se_3Te$ | $P3m1$ | $\beta_2$ | 4.00 | 1.20 | 1.05 | 0.00006 | 0.0045 | 0.030 |
| | $CaAl_2Se_3Te$ | $P3m1$ | $\beta_2$ | 4.09 | 0.82 | 0.68 | 0.00012 | 0.0045 | 0.054 |
| | $SrAl_2Se_3Te$ | $P3m1$ | $\beta_2$ | 4.14 | 0.69 | 0.55 | 0.00010 | 0.0044 | 0.047 |
| | $MgAl_2Te_3S$ | $P3m1$ | $\beta_2$ | 4.07 | 0.00 | 0.02 | \ | \ | \ |
| | $CaAl_2Te_3S$ | $P3m1$ | $\beta_2$ | 4.12 | 0.00 | 0.05 | \ | \ | \ |
| | $SrAl_2Te_3S$ | $P3m1$ | $\beta_2$ | 4.20 | 0.00 | 0.09 | \ | \ | \ |
| | $MgAl_2Te_3Se$ | $P3m1$ | $\beta_2$ | 4.13 | 0.00 | 0.05 | \ | \ | \ |
| | $CaAl_2Te_3Se$ | $P3m1$ | $\beta_2$ | 4.19 | 0.00 | 0.01 | \ | \ | \ |
| | $SrAl_2Te_3Se$ | $P3m1$ | $\beta_2$ | 4.23 | 0.00 | 0.02 | \ | \ | \ |
| $MGa_2Z_3Z'$ | $MgGa_2S_3Se$ | $P3m1$ | $\beta_2$ | 3.77 | 1.33 | 1.29 | \ | \ | \ |
| | $CaGa_2S_3Se$ | $P3m1$ | $\beta_2$ | 3.87 | 0.82 | 0.78 | \ | \ | \ |
| | $SrGa_2S_3Se$ | $P3m1$ | $\beta_2$ | 3.93 | 0.65 | 0.61 | \ | \ | \ |
| | $MgGa_2S_3Te$ | $P3m1$ | $\beta_2$ | 3.88 | 0.46 | 0.20 | \ | \ | \ |
| | $CaGa_2S_3Te$ | $P3m1$ | $\beta_2$ | 3.99 | 0.35 | 0.14 | \ | \ | \ |
| | $SrGa_2S_3Te$ | $P3m1$ | $\beta_2$ | 4.05 | 0.26 | 0.09 | \ | \ | \ |
| | $MgGa_2Se_3S$ | $P3m1$ | $\beta_2$ | 3.86 | 0.34 | 0.21 | 0.00038 | 0.0047 | 0.159 |
| | $CaGa_2Se_3S$ | $P3m1$ | $\beta_2$ | 3.93 | 0.00 | 0.04 | \ | \ | \ |
| | $SrGa_2Se_3S$ | $P3m1$ | $\beta_2$ | 3.99 | 0.00 | 0.04 | \ | \ | \ |
| | $MgGa_2Se_3Te$ | $P3m1$ | $\beta_2$ | 4.02 | 0.49 | 0.32 | 0.00017 | 0.0045 | 0.073 |
| | $CaGa_2Se_3Te$ | $P3m1$ | $\beta_2$ | 4.11 | 0.10 | 0.00 | \ | \ | \ |
| | $SrGa_2Se_3Te$ | $P3m1$ | $\beta_2$ | 4.17 | 0.00 | 0.05 | \ | \ | \ |
| | $MgGa_2Te_3S$ | $P3m1$ | $\beta_2$ | 4.10 | 0.00 | 0.00 | \ | \ | \ |
| | $CaGa_2Te_3S$ | $P3m1$ | $\beta_2$ | 4.15 | 0.00 | 0.10 | \ | \ | \ |
| | $SrGa_2Te_3S$ | $P3m1$ | $\beta_2$ | 4.20 | 0.00 | 0.09 | \ | \ | \ |
| | $MgGa_2Te_3Se$ | $P3m1$ | $\beta_2$ | 4.16 | 0.00 | 0.00 | \ | \ | \ |
| | $CaGa_2Te_3Se$ | $P3m1$ | $\beta_2$ | 4.22 | 0.00 | 0.05 | \ | \ | \ |
| | $SrGa_2Te_3Se$ | $P3m1$ | $\beta_2$ | 4.26 | 0.00 | 0.01 | 0.00398 | 0.0128 | 0.620 |

**Table S3.** Lattice parameters, bandgaps and Rashba parameters of "5-Janus" configuration in 2D $MAA'Z_2Z'_2$ systems.

| Classes | Systems | Structure | Phase | $a=b$ (Å) | $E_{g\text{-PBE}}$ (eV) | $E_{g\text{-PBE-SOC}}$ (eV) | $E_R$ (eV) | $k_R$ ($Å^{-1}$) | $\alpha_R$ (eVÅ) |
|---|---|---|---|---|---|---|---|---|---|
| $MAlGaZ_3Z'$ | $MgAlGaS_3Se$ | $P3m1$ | $\beta_2$ | 3.76 | 1.87 | 1.83 | 0.00079 | 0.0049 | 0.326 |
| | $CaAlGaS_3Se$ | $P3m1$ | $\beta_2$ | 3.85 | 1.36 | 1.32 | 0.00094 | 0.0047 | 0.397 |
| | $SrAlGaS_3Se$ | $P3m1$ | $\beta_2$ | 3.91 | 1.15 | 1.11 | 0.00100 | 0.0047 | 0.430 |
| | $MgAlGaS_3Te$ | $P3m1$ | $\beta_2$ | 3.86 | 1.13 | 0.85 | \ | \ | \ |
| | $CaAl_2S_3Te$ | $P3m1$ | $\beta_2$ | 3.97 | 1.07 | 0.83 | \ | \ | \ |
| | $SrAlGaS_3Te$ | $P3m1$ | $\beta_2$ | 4.03 | 1.00 | 0.78 | \ | \ | \ |
| | $MgAlGaSe_3S$ | $P3m1$ | $\beta_2$ | 3.84 | 0.81 | 0.67 | 0.00066 | 0.0047 | 0.278 |
| | $CaAlGaSe_3S$ | $P3m1$ | $\beta_2$ | 3.91 | 0.39 | 0.26 | 0.00080 | 0.0047 | 0.342 |
| | $SrAlGaSe_3S$ | $P3m1$ | $\beta_2$ | 3.96 | 0.24 | 0.11 | 0.00082 | 0.0046 | 0.357 |
| | $MgAlGaSe_3Te$ | $P3m1$ | $\beta_2$ | 4.00 | 1.07 | 0.88 | \ | \ | \ |
| | $CaAlGaSe_3Te$ | $P3m1$ | $\beta_2$ | 4.09 | 0.73 | 0.58 | \ | \ | \ |
| | $SrAlGaSe_3Te$ | $P3m1$ | $\beta_2$ | 4.15 | 0.61 | 0.46 | \ | \ | \ |
| | $MgAlGaTe_3S$ | $P3m1$ | $\beta_2$ | 4.08 | 0.00 | 0.03 | \ | \ | \ |
| | $CaAlGaTe_3S$ | $P3m1$ | $\beta_2$ | 4.13 | 0.00 | 0.04 | \ | \ | \ |
| | $SrAlGaTe_3S$ | $P3m1$ | $\beta_2$ | 4.17 | 0.00 | 0.04 | \ | \ | \ |
| | $MgAlGaTe_3Se$ | $P3m1$ | $\beta_2$ | 4.14 | 0.00 | 0.06 | \ | \ | \ |
| | $CaAlGaTe_3Se$ | $P3m1$ | $\beta_2$ | 4.20 | 0.00 | 0.00 | \ | \ | \ |
| | $SrAlGaTe_3Se$ | $P3m1$ | $\beta_2$ | 4.24 | 0.00 | 0.00 | \ | \ | \ |
| $MGaAlZ_3Z'$ | $MgGaAlS_3Se$ | $P3m1$ | $\beta_2$ | 3.76 | 1.42 | 1.39 | \ | \ | \ |
| | $CaGaAlS_3Se$ | $P3m1$ | $\beta_2$ | 3.86 | 0.92 | 0.88 | \ | \ | \ |
| | $SrGaAlS_3Se$ | $P3m1$ | $\beta_2$ | 3.92 | 0.73 | 0.70 | \ | \ | \ |
| | $MgGaAlS_3Te$ | $P3m1$ | $\beta_2$ | 3.87 | 0.65 | 0.38 | \ | \ | \ |
| | $CaGaAlS_3Te$ | $P3m1$ | $\beta_2$ | 3.98 | 0.49 | 0.28 | \ | \ | \ |
| | $SrGaAlS_3Te$ | $P3m1$ | $\beta_2$ | 4.05 | 0.39 | 0.21 | \ | \ | \ |

| | | | | | | | | |
|---|---|---|---|---|---|---|---|---|
| MgGaAlSe$_3$S | *P*3*m*1 | $\beta_2$ | 3.84 | 0.42 | 0.30 | 0.00008 | 0.0048 | 0.035 |
| CaGaAlSe$_3$S | *P*3*m*1 | $\beta_2$ | 3.92 | 0.00 | 0.03 | \ | \ | \ |
| SrGaAlSe$_3$S | *P*3*m*1 | $\beta_2$ | 3.97 | 0.00 | 0.03 | \ | \ | \ |
| MgGaAlSe$_3$Te | *P*3*m*1 | $\beta_2$ | 4.01 | 0.59 | 0.45 | 0.00082 | 0.0045 | 0.181 |
| CaGaAlSe$_3$Te | *P*3*m*1 | $\beta_2$ | 4.10 | 0.17 | 0.05 | 0.00018 | 0.0044 | 0.080 |
| SrGaAlSe$_3$Te | *P*3*m*1 | $\beta_2$ | 4.16 | 0.00 | 0.03 | \ | \ | \ |
| MgGaAlTe$_3$S | *P*3*m*1 | $\beta_2$ | 4.10 | 0.00 | 0.00 | \ | \ | \ |
| CaGaAlTe$_3$S | *P*3*m*1 | $\beta_2$ | 4.15 | 0.00 | 0.11 | \ | \ | \ |
| SrGaAlTe$_3$S | *P*3*m*1 | $\beta_2$ | 4.19 | 0.00 | 0.11 | \ | \ | \ |
| MgGaAlTe$_3$Se | *P*3*m*1 | $\beta_2$ | 4.15 | 0.00 | 0.00 | \ | \ | \ |
| CaGaAlTe$_3$Se | *P*3*m*1 | $\beta_2$ | 4.21 | 0.00 | 0.06 | \ | \ | \ |
| SrGaAlTe$_3$Se | *P*3*m*1 | $\beta_2$ | 4.25 | 0.00 | 0.06 | \ | \ | \ |

**Table S4.** Lattice parameters, bandgaps and Rashba parameters of "3-Janus" configuration in 2D MAA'Z$_2$Z'$_2$ systems.

| Classes | Systems | Structure | Phase | *a*=*b* (Å) | $E_{g\text{-PBE}}$ (eV) | $E_{g\text{-PBE-SOC}}$ (eV) | $E_R$ (eV) | $k_R$ (Å$^{-1}$) | $\alpha_R$ (eVÅ) |
|---|---|---|---|---|---|---|---|---|---|
| | MgAlGaS$_4$ | *P*3*m*1 | $\beta_2$ | 3.70 | 1.57 | 1.53 | \ | \ | \ |
| | CaAlGaS$_4$ | *P*3*m*1 | $\beta_2$ | 3.79 | 1.06 | 1.03 | \ | \ | \ |
| | SrAlGaS$_4$ | *P*3*m*1 | $\beta_2$ | 3.85 | 0.87 | 0.83 | \ | \ | \ |
| | MgAlGaSe$_4$ | *P*3*m*1 | $\beta_2$ | 3.90 | 0.78 | 0.65 | 0.00060 | 0.0054 | 0.223 |
| MAlGaZ$_4$ | CaAlGaSe$_4$ | *P*3*m*1 | $\beta_2$ | 3.98 | 0.33 | 0.21 | 0.00070 | 0.0053 | 0.265 |
| | SrAlGaSe$_4$ | *P*3*m*1 | $\beta_2$ | 4.04 | 0.19 | 0.07 | 0.00057 | 0.0052 | 0.220 |
| | MgAlGaTe$_4$ | *P*3*m*1 | $\beta_2$ | 4.24 | 0.30 | 0.08 | 0.00443 | 0.0099 | 0.894 |
| | CaAlGaTe$_4$ | *P*3*m*1 | $\beta_2$ | 4.31 | 0.00 | 0.02 | \ | \ | \ |
| | SrAlGaTe$_4$ | *P*3*m*1 | $\beta_2$ | 4.36 | 0.00 | 0.02 | \ | \ | \ |
| | MgAlInS$_4$ | *P*3*m*1 | $\beta_2$ | 3.79 | 1.67 | 1.64 | \ | \ | \ |
| | CaAlInS$_4$ | *P*3*m*1 | $\beta_2$ | 3.89 | 1.23 | 1.19 | \ | \ | \ |
| | SrAlInS$_4$ | *P*3*m*1 | $\beta_2$ | 3.96 | 1.05 | 1.02 | \ | \ | \ |
| | MgAlInSe$_4$ | *P*3*m*1 | $\beta_2$ | 3.98 | 1.02 | 0.89 | \ | \ | \ |
| MAlInZ$_4$ | CaAlInSe$_4$ | *P*3*m*1 | $\beta_2$ | 4.08 | 0.62 | 0.49 | 0.00005 | 0.0045 | 0.025 |
| | SrAlInSe$_4$ | *P*3*m*1 | $\beta_2$ | 4.13 | 0.48 | 0.35 | 0.00014 | 0.0044 | 0.062 |
| | MgAlInTe$_4$ | *P*3*m*1 | $\beta_2$ | 4.32 | 0.65 | 0.43 | 0.00028 | 0.0042 | 0.131 |
| | CaAlInTe$_4$ | *P*3*m*1 | $\beta_2$ | 4.39 | 0.25 | 0.00 | \ | \ | \ |
| | SrAlInTe$_4$ | *P*3*m*1 | $\beta_2$ | 4.45 | 0.13 | 0.05 | \ | \ | \ |
| | MgGaInS$_4$ | *P*3*m*1 | $\beta_2$ | 3.81 | 1.21 | 1.18 | \ | \ | \ |
| | CaGaInS$_4$ | *P*3*m*1 | $\beta_2$ | 3.91 | 0.72 | 0.69 | \ | \ | \ |
| | SrGaInS$_4$ | *P*3*m*1 | $\beta_2$ | 3.98 | 0.55 | 0.52 | \ | \ | \ |
| | MgGaInSe$_4$ | *P*3*m*1 | $\beta_2$ | 4.00 | 0.54 | 0.42 | 0.00037 | 0.0046 | 0.164 |
| MGaInZ$_4$ | CaGaInSe$_4$ | *P*3*m*1 | $\beta_2$ | 4.09 | 0.11 | 0.00 | \ | \ | \ |
| | SrGaInSe$_4$ | *P*3*m*1 | $\beta_2$ | 4.16 | 0.00 | 0.05 | \ | \ | \ |
| | MgGaInTe$_4$ | *P*3*m*1 | $\beta_2$ | 4.32 | 0.14 | 0.03 | \ | \ | \ |
| | CaGaInTe$_4$ | *P*3*m*1 | $\beta_2$ | 4.41 | 0.00 | 0.04 | \ | \ | \ |
| | SrGaInTe$_4$ | *P*3*m*1 | $\beta_2$ | 4.46 | 0.00 | 0.03 | \ | \ | \ |

| Classes | Systems | Structure | Phase | *a*=*b* (Å) | $E_{g\text{-PBE}}$ (eV) | $E_{g\text{-PBE-SOC}}$ (eV) | $E_R$ (eV) | $k_R$ (Å$^{-1}$) | $\alpha_R$ (eVÅ) |
|---|---|---|---|---|---|---|---|---|---|
| | MgAl$_2$S$_2$Se$_2$ | *P*3*m*1 | $\beta_2$ | 3.79 | 1.15 | 1.02 | \ | \ | \ |
| | CaAl$_2$S$_2$Se$_2$ | *P*3*m*1 | $\beta_2$ | 3.87 | 0.94 | 0.82 | \ | \ | \ |
| MAl$_2$Z$_2$ Z'$_2$ | SrAl$_2$S$_2$Se$_2$ | *P*3*m*1 | $\beta_2$ | 3.93 | 0.79 | 0.67 | \ | \ | \ |
| | MgAl$_2$S$_2$Te$_2$ | *P*3*m*1 | $\beta_2$ | 3.96 | 0.00 | 0.03 | \ | \ | \ |
| | CaAl$_2$S$_2$Te$_2$ | *P*3*m*1 | $\beta_2$ | 4.04 | 0.00 | 0.05 | \ | \ | \ |
| | SrAl$_2$S$_2$Te$_2$ | *P*3*m*1 | $\beta_2$ | 4.10 | 0.00 | 0.02 | \ | \ | \ |
| | $MgAl_2S_2eTe_2$ | $P3m1$ | $\beta_2$ | 4.06 | 0.00 | 0.06 | \ | \ | \ |
| | $CaAl_2S_2eTe_2$ | $P3m1$ | $\beta_2$ | 4.14 | 0.00 | 0.05 | \ | \ | \ |
| | $SrAl_2S_2eTe_2$ | $P3m1$ | $\beta_2$ | 4.19 | 0.00 | 0.04 | \ | \ | \ |
| $MGa_2Z_2$ $Z'_2$ | $MgGa_2S_2Se_2$ | $P3m1$ | $\beta_2$ | 3.81 | 0.38 | 0.26 | \ | \ | \ |
| | $CaGa_2S_2Se_2$ | $P3m1$ | $\beta_2$ | 3.90 | 0.10 | 0.00 | \ | \ | \ |
| | $SrGa_2S_2Se_2$ | $P3m1$ | $\beta_2$ | 3.96 | 0.00 | 0.04 | \ | \ | \ |
| | $MgGa_2S_2Te_2$ | $P3m1$ | $\beta_2$ | 3.98 | 0.00 | 0.02 | \ | \ | \ |
| | $CaGa_2S_2Te_2$ | $P3m1$ | $\beta_2$ | 4.07 | 0.00 | 0.01 | \ | \ | \ |
| | $SrGa_2S_2Te_2$ | $P3m1$ | $\beta_2$ | 4.13 | 0.00 | 0.00 | \ | \ | \ |
| | $MgGa_2S_2eTe_2$ | $P3m1$ | $\beta_2$ | 4.08 | 0.00 | 0.06 | \ | \ | \ |
| | $CaGa_2S_2eTe_2$ | $P3m1$ | $\beta_2$ | 4.17 | 0.00 | 0.03 | \ | \ | \ |
| | $SrGa_2S_2eTe_2$ | $P3m1$ | $\beta_2$ | 4.22 | 0.00 | 0.02 | \ | \ | \ |

| Classes | Systems | Structure | Phase | $a=b$ (Å) | $E_{g\text{-PBE}}$ (eV) | $E_{g\text{-PBE-SOC}}$ (eV) | $E_R$ (eV) | $k_R$ (Å$^{-1}$) | $\alpha_R$ (eVÅ) |
|---|---|---|---|---|---|---|---|---|---|
| MAlGa $Z_2Z'_2$ | $MgAlGaS_2Se_2$ | $P3m1$ | $\beta_2$ | 3.80 | 0.97 | 0.84 | 0.00036 | 0.0047 | 0.153 |
| | $CaAlGaS_2Se_2$ | $P3m1$ | $\beta_2$ | 3.88 | 0.68 | 0.56 | 0.00072 | 0.0047 | 0.307 |
| | $SrAlGaS_2Se_2$ | $P3m1$ | $\beta_2$ | 3.94 | 0.50 | 0.38 | 0.00083 | 0.0046 | 0.359 |
| | $MgAlGaS_2Te_2$ | $P3m1$ | $\beta_2$ | 3.97 | 0.00 | 0.05 | \ | \ | \ |
| | $CaAlGaS_2Te_2$ | $P3m1$ | $\beta_2$ | 4.05 | 0.00 | 0.03 | \ | \ | \ |
| | $SrAlGaS_2Te_2$ | $P3m1$ | $\beta_2$ | 4.10 | 0.00 | 0.02 | \ | \ | \ |
| | $MgAlGaSe_2Te_2$ | $P3m1$ | $\beta_2$ | 4.07 | 0.00 | 0.07 | \ | \ | \ |
| | $CaAlGaSe_2Te_2$ | $P3m1$ | $\beta_2$ | 4.14 | 0.00 | 0.05 | \ | \ | \ |
| | $SrAlGaSe_2Te_2$ | $P3m1$ | $\beta_2$ | 4.20 | 0.00 | 0.04 | \ | \ | \ |
| MGaAl $Z_2Z'_2$ | $MgGaAlS_2Se_2$ | $P3m1$ | $\beta_2$ | 3.80 | 0.47 | 0.34 | \ | \ | \ |
| | $CaGaAlS_2Se_2$ | $P3m1$ | $\beta_2$ | 3.89 | 0.18 | 0.08 | \ | \ | \ |
| | $SrGaAlS_2Se_2$ | $P3m1$ | $\beta_2$ | 3.94 | 0.00 | 0.00 | \ | \ | \ |
| | $MgGaAlS_2Te_2$ | $P3m1$ | $\beta_2$ | 3.98 | 0.00 | 0.02 | \ | \ | \ |
| | $CaGaAlS_2Te_2$ | $P3m1$ | $\beta_2$ | 4.07 | 0.00 | 0.01 | \ | \ | \ |
| | $SrGaAlS_2Te_2$ | $P3m1$ | $\beta_2$ | 4.12 | 0.00 | 0.00 | \ | \ | \ |
| | $MgGaAlSe_2Te_2$ | $P3m1$ | $\beta_2$ | 4.08 | 0.00 | 0.05 | \ | \ | \ |
| | $CaGaAlSe_2Te_2$ | $P3m1$ | $\beta_2$ | 4.16 | 0.00 | 0.03 | \ | \ | \ |
| | $SrGaAlSe_2Te_2$ | $P3m1$ | $\beta_2$ | 4.21 | 0.00 | 0.02 | \ | \ | \ |

**Table S5**: The energy of $MgAl_2S_4$ monolayer with different phases.

| Phase | $\alpha_1$ | $\alpha_2$ | $\beta_1$ | $\beta_2$ |
|---|---|---|---|---|
| Energy (eV) | -34.45438409 | -34.43282507 | -34.82940267 | -34.84206603 |

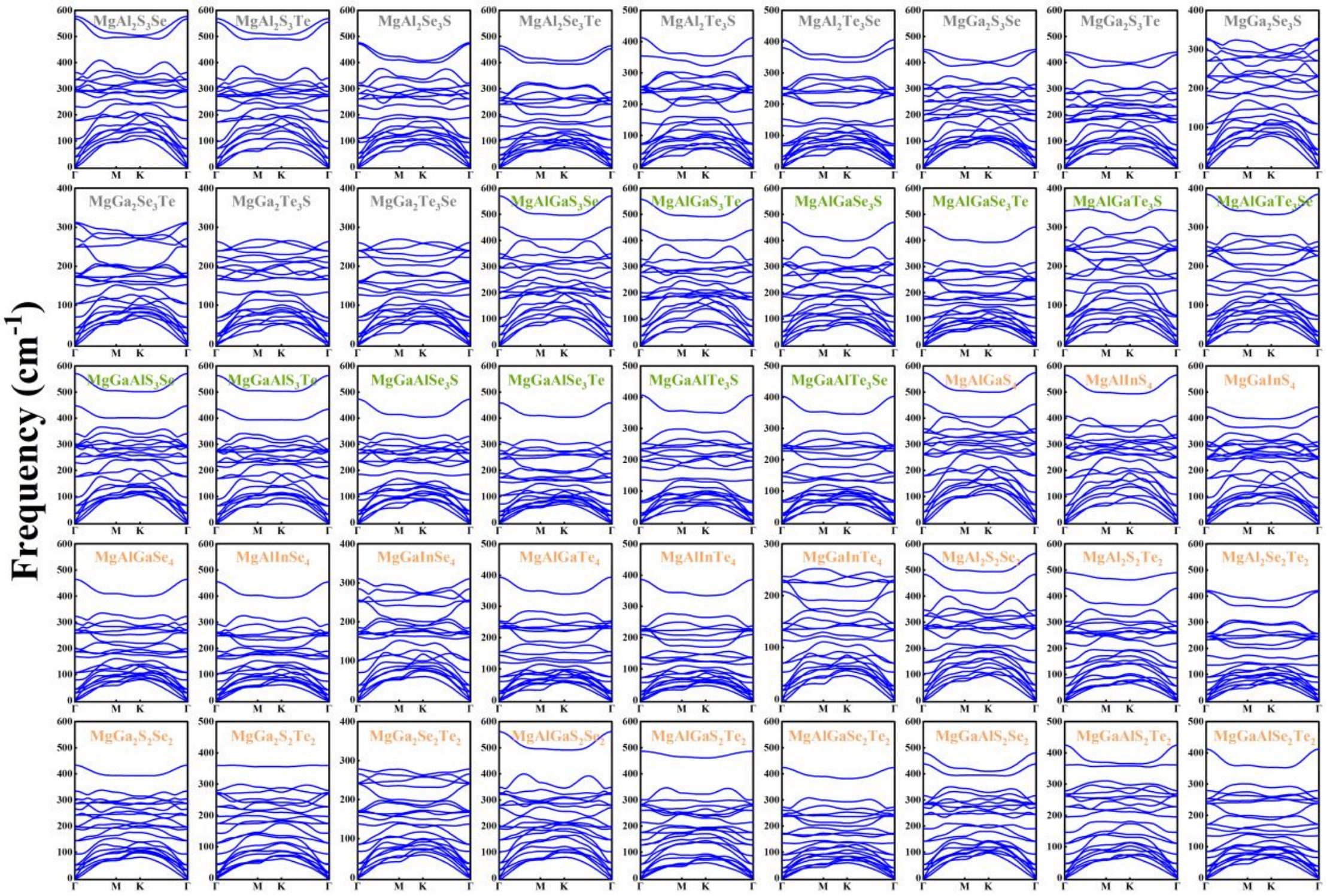

**Figure S1**. Phonon spectrum of all MgAA'$Z_xZ'_{(4-x)}$ monolayers.

**Table S6**: Structural properties of $MgAl_2S_4$ monolayer and seven monolayers with different '7/5/3-janus' configurations. *a* is the Lattice parameter. Z/A denotes the sequence of Z atoms and A atoms in POSCAR, while $Z_{Z1}$, $Z_{Z2}$, $Z_{Z3}$, $Z_{Z4}$, $Z_M$, $Z_{A1}$ and $Z_{A2}$ represent the *z*-component of their atomic positions, all given in fractional coordinates.

| $\beta_2$ phase | Z/A | *a* (Å) | $Z_{Z1}$ | $Z_{Z2}$ | $Z_{Z3}$ | $Z_{Z4}$ | $Z_M$ | $Z_{A1}$ | $Z_{A2}$ |
|---|---|---|---|---|---|---|---|---|---|
| $MgAl_2S_4$ | SSSS<br>AlAl | 3.689 | 0.170 | 0.464 | 0.272 | 0.362 | 0.317 | 0.432 | 0.202 |
| $MgAl_2S_3Se$ | SeSSS<br>AlAl | 3.749 | 0.166 | 0.462 | 0.274 | 0.361 | 0.317 | 0.432 | 0.203 |
| $MgAlGaS_3Se$ | SeSSS<br>AlGa | 3.760 | 0.165 | 0.462 | 0.274 | 0.361 | 0.317 | 0.432 | 0.203 |
| $MgGaAlS_3Se$ | SSSSe<br>AlGa | 3.762 | 0.169 | 0.273 | 0.360 | 0.467 | 0.316 | 0.430 | 0.201 |
| $MgAlGaS_4$ | SSSS<br>AlGa | 3.70 | 0.169 | 0.463 | 0.273 | 0.361 | 0.317 | 0.431 | 0.202 |
| $MgAl_2S_2Se_2$ | SeSeSS<br>AlAl | 3.787 | 0.162 | 0.273 | 0.364 | 0.464 | 0.321 | 0.434 | 0.198 |
| $MgAlGaS_2Se_2$ | SeSeSS<br>AlGa | 3.799 | 0.161 | 0.374 | 0.464 | 0.364 | 0.321 | 0.434 | 0.198 |
| $MgGaAlS_2Se_2$ | SSSeSe | 3.801 | 0.167 | 0.270 | 0.471 | 0.360 | 0.313 | 0.435 | 0.199 |

AlGa

An example of the structural file (POSCAR) for $\beta_2$-$MgAl_2S_4$ was provided:

$\beta_2$-$MgAl_2S_4$

1

$a$ 0.000 0.000

-0.5$a$ $\sqrt{3}a/2$ 0.000

0.000 0.000 30.000

$Z_1$ $Z_2$ $Z_3$ $Z_4$ M $A_1$ $A_2$

1 1 1 1 1 1 1

Direct

0.333 0.666 $Z_{Z1}$

0.666 0.333 $Z_{Z2}$

0.666 0.333 $Z_{Z3}$

0.333 0.666 $Z_{Z4}$

0.000 0.000 $Z_M$

0.333 0.666 $Z_{A1}$

0.666 0.333 $Z_{A2}$

## C. Band structures of monolayer $MA_2Z_4$ and Janus $MAA'Z_xZ'_{(4-x)}$ TIs

This section provides the band structures of monolayer $MA_2Z_4$ and Janus $MAA'Z_xZ'_{(4-x)}$ topological insulators (TIs), as well as the orbital projection bands of $CaAA'Z_4$ and $SrAA'Z_4$ monolayers. Replacing the central atom M from Mg to Ca or Sr reduces the bandgaps of $MAA'Z_4$ monolayers to 0–1.06 eV ($CaAA'Z_4$) and 0–0.87 eV ($SrAA'Z_4$), respectively. Moreover, the bandgaps of the $MAA'Z_4$ (M=Ca, Sr) monolayers also gradually decrease with the doping of A and Z atoms, and the CBM and VBM mainly derived from their respective A-*s* and Z-*p* orbitals, as shown in Figure S5, similar to that of $MgAA'Z_4$. The band structures and properties of $CaAA'Z_4$ and $SrAA'Z_4$ monolayers are similar to those of $MgAA'Z_4$, suggesting that the electronic structure principles derived from Mg-centered monolayers are also applicable to other $MAA'Z_xZ'_{(4-x)}$ monolayers. The distinction lies in that when M is Ca or Sr, due to the heavier central atom, the overall bandgap of the $MAA'Z_xZ'_{(4-x)}$ monolayers is smaller, facilitating band inversion phenomena and thus making the transition from NI to TI more accessible.

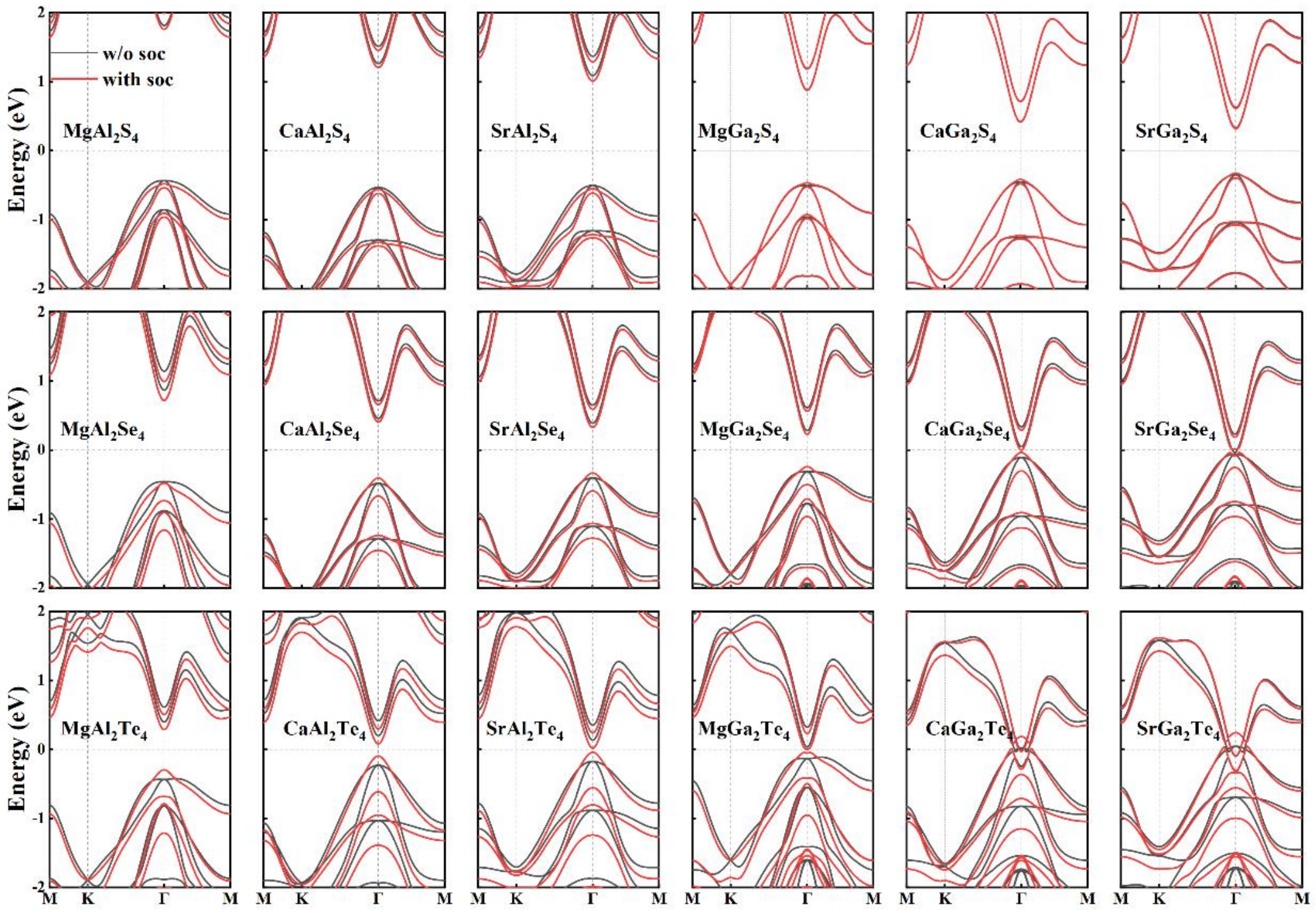

**Figure S2**. Band structures of monolayer $MA_2Z_4$.

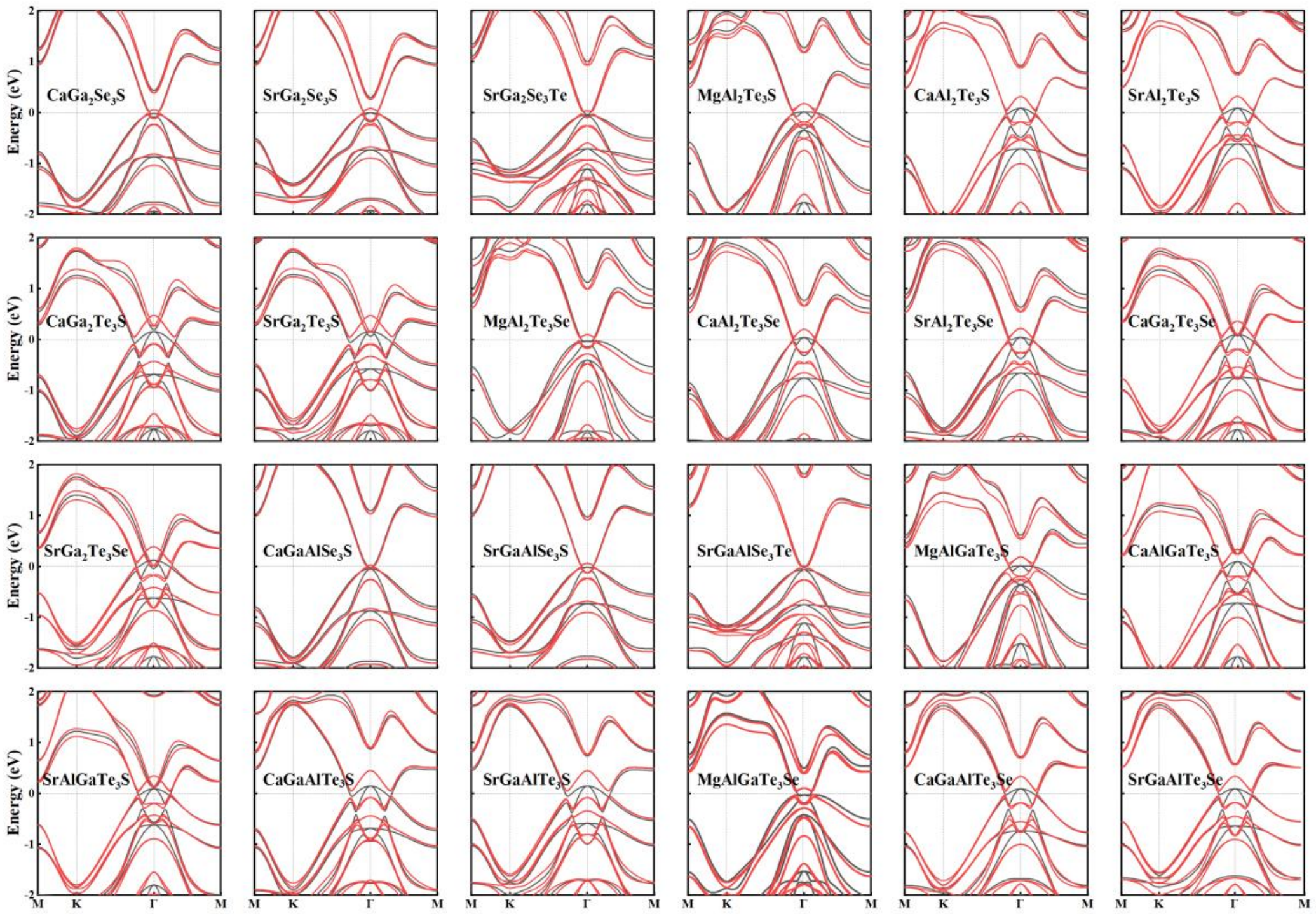

**Figure S3**. Band structures of '7/5-Janus' $MAA'Z_xZ'_{(4-x)}$ TIs.

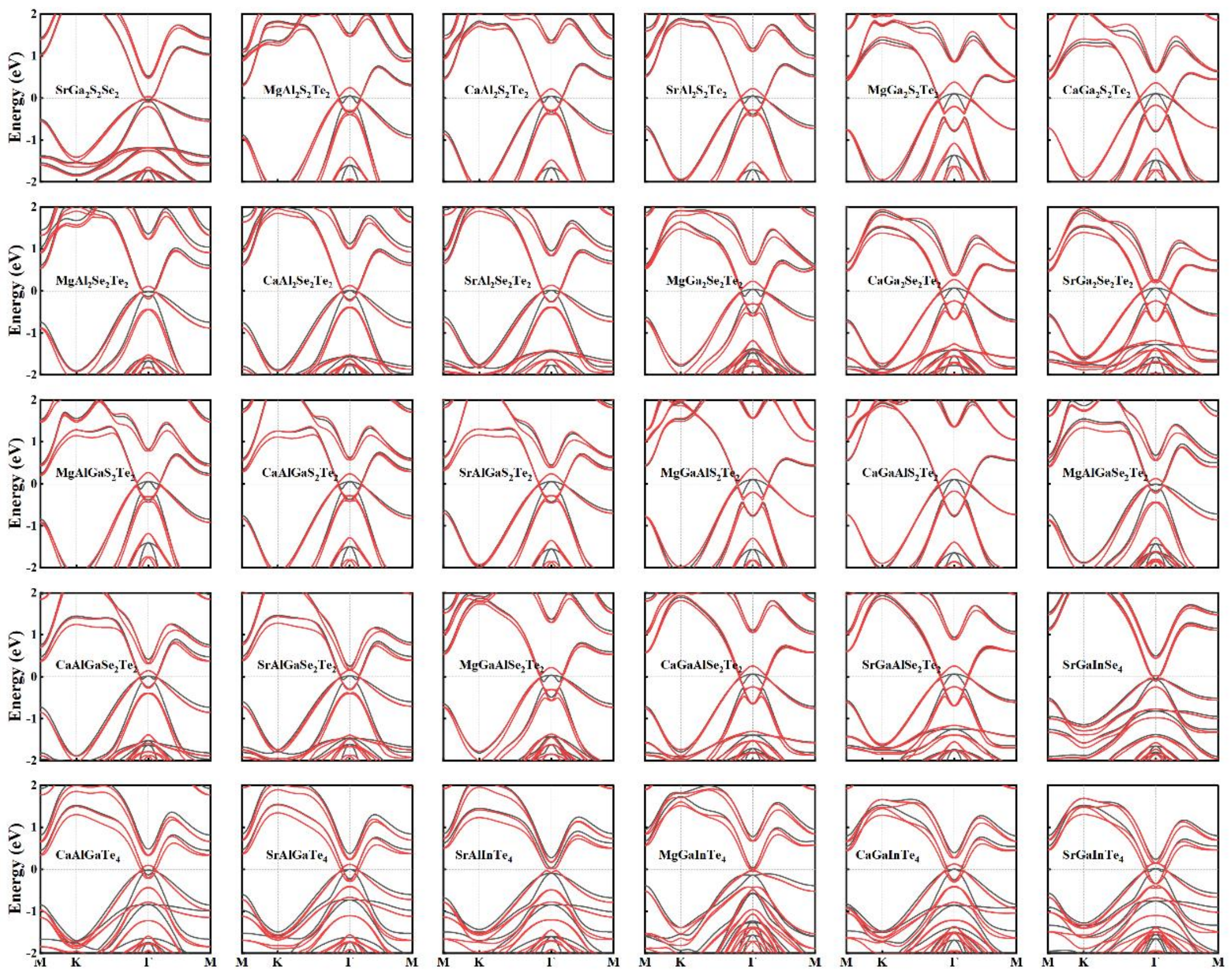

**Figure S4**. Band structures of '3-Janus' $MAA'Z_xZ'_{(4-x)}$ TIs.

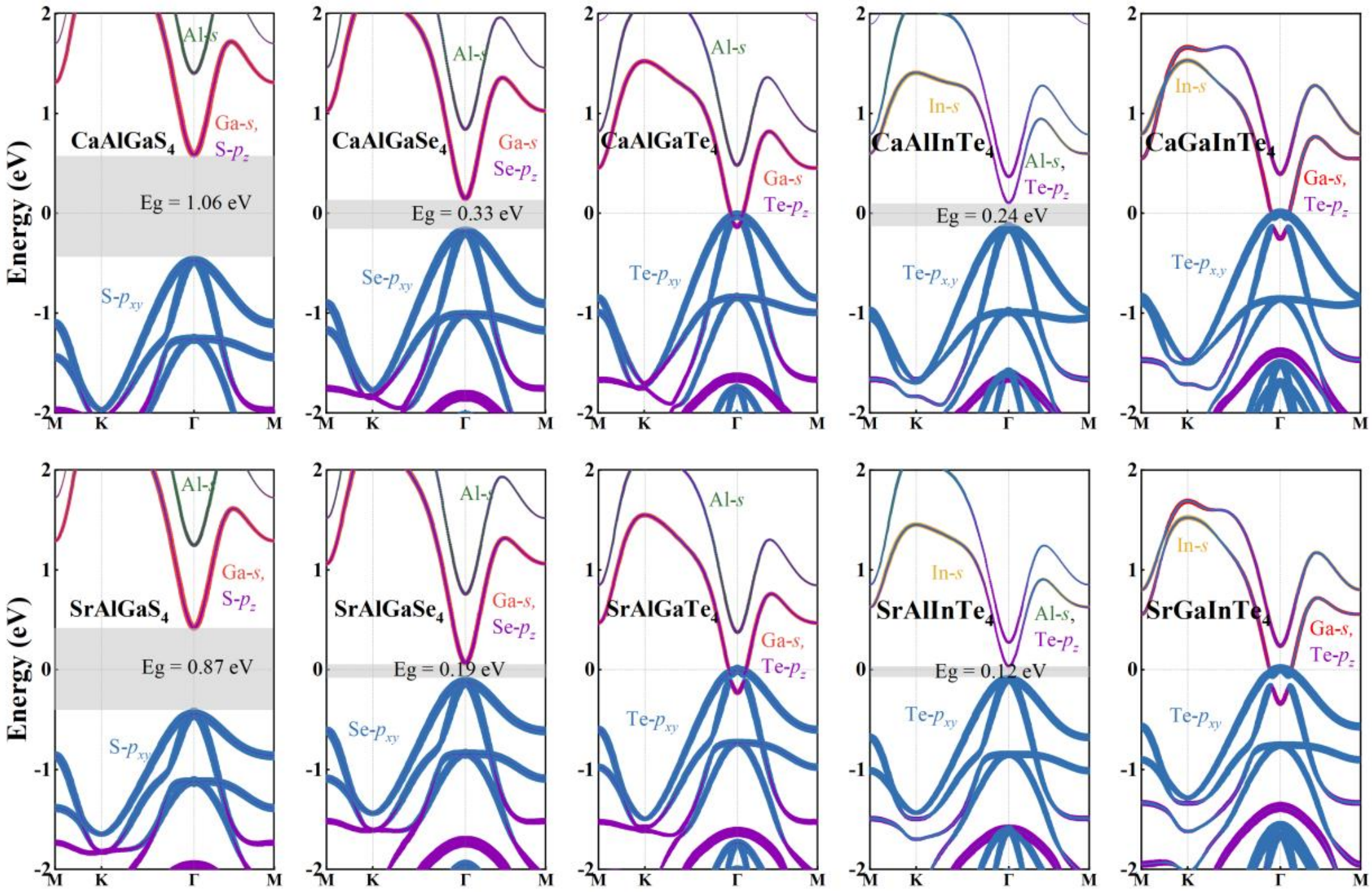

**Figure S5**. Orbital projected band structures of (a) $CaAA'Z_4$ monolayers and (b) $SrAA'Z_4$ monolayers without considering SOC.

## D. Analysis of Rashba effect and topological states in 2D MAA'$Z_xZ'_{(4-x)}$ family

The Rashba effect in MAA'$Z_xZ'_{(4-x)}$ monolayers is analyzed via intrinsic electric field and charge transfer, and the orbital and spin texture changes of Janus MAA'$Z_xZ'_{(4-x)}$ monolayers under applied strain are studied in detail.

To quantitatively analyze the Rashba effect of Janus MAA'$Z_xZ'_{(4-x)}$ monolayers, we further explore the underlying mechanism of the Rashba effect from the perspective of intrinsic electric field ($E_{in}$) and SOC strength. The $E_{in}$ is positively correlated to the charge transfer in Janus monolayers, as reported in our previous work [11]. The diagram of the intrinsic electric field associated with charge transfer in Janus MAA'$Z_xZ'_{(4-x)}$ monolayers shows that electrons can be transferred from the M, A, and A' layers to the adjacent Z and Z' layers, forming a localized electric field between adjacent atomic layers, and that the sum of charge transfers ($Q$) can reflect the strength of $E_{in}$. On the other hand, the charge transfer could be visually reflected from the charge density difference in real space. For the $MA_2Z_4$ monolayers, the total $Q = (e_1^- + e_3^- + e_5^-) - (e_2^- + e_4^- + e_6^-) = 0$, which corresponds to the $E_{in} = (E_1 + E_3 + E_5) - (E_2 + E_4 + E_6) = 0$. For Janus MAA'$Z_xZ'_{(4-x)}$ monolayers, due to the difference in electronegativity and atomic size of the elements on both sides of the central atom M, the charge transfer behavior will change, resulting in unequal transfer of charges in the upward and downward directions and local electric fields, with a nonzero total $Q$ and $E_{in}$. For example, from the calculated Bader charges of $MgAl_2S_4$ and $MgAlGaTe_4$ monolayers (Table S7), we can observe that the atoms on both sides of $MgAl_2S_4$ monolayer carry nearly identical charges, and $Q$ is almost 0 *e*, while the $Q$ of $MgAlGaTe_4$ monolayer is 0.777 *e*. The analysis of the relationship between $Q$ and $\alpha_R$ of MgAA'$Z_xZ'_{(4-x)}$ monolayers with ideal Rashba effect shows that materials with larger Rashba constants tend to exhibit relatively larger $Q$. From the most fundamental perspective, the SOC strength of the system is also a key factor that determines the $\alpha_R$ [12]. The SOC can be reflected by atomic number, which has been verified by the fact that the SOC strength of transition metal dichalcogenide monolayers increases with the increase of chalcogen atomic number [13]. Taking $MgAlGaSe_4$ and $MgAlGaTe_4$ as examples, despite their comparable $Q$, the $MgAlGaTe_4$ monolayer exhibits significantly larger $\alpha_R$ due to the presence of Te, which has a higher atomic number. In summary, due to the large SOC and the existence of $E_{in}$, the atomic number and charge transfer are key factors influencing the strength of the Rashba effect. Furthermore, the orbital contribution also indicates that the Rashba splitting bands all have contributions from the Z-$p_z$ orbital.

In MAA'Z$_x$Z'$_{(4-x)}$ Rashba systems, an intrinsic electric field $E_z$ breaks the inversion symmetry along the $z$-direction, resulting in Rashba splitting of the bands dominated by $p_z$ orbitals, as the interaction between $E_z$ and atomic orbitals is governed by their parity symmetry. The $E_z$ potential energy term ($V(z) =- eE_z z$) have odd parity under $z$ reflection. Due to the broken inversion symmetry, the parity of wave function of the $p_z$ orbital is not strictly odd ($p_z(-z) \neq - p_z(z)$), but a linear combination containing both odd and even components ($p_z(z) = p_e(z) + p_o(z)$), which is also applicable in previous studies [14–16]. Therefore, the matrix element $\langle p_z|z|p_z\rangle$, which involves an integral over the asymmetric distribution of $p_z$ along $z$, contains even, resulting in a *non-zero* integral:

$$\langle p_z|z|p_z\rangle = \int p_z^*(z) \cdot z \cdot p_z(z)dz \neq 0$$

However, the $p_{x,\,y}$ orbitals are oriented *in-plane* ($x$ and $y$ directions) and exhibit even parity under mirror reflection in the *z-plane*. Their product is odd, leading to an integral that vanishes over symmetric limits:

$$\langle p_{x,y}|z|p_{x,y}\rangle = \int p_{x,y}^*(z) \cdot z \cdot p_{x,y}(z)dz = 0$$

For the $p_z$ orbital, its asymmetric distribution along the $z$ enables it to "sense" the vertical field gradient, breaking inversion symmetry and realizing Rashba splitting. In contrast, the vertical field cannot disturb the energy degeneracy of $p_{x,\,y}$ orbitals due to their planar symmetry.

When biaxial strain is applied to $MgGaInTe_4$ TIs within the range of -2% to 2%, a topological phase transition occurs, accompanied by modifications in the spin texture of the Rashba-split bands. In detail, the band inversion disappears at -2% strain and shows a semiconductor nature with a bandgap (0.03 eV), which maintains a normal Rashba spin splitting at CBM with switching on SOC, similar to that observed in $MgAlGaTe_4$ monolayer. The $Z2$ invariant is found to be 0 (as shown in Figure S9), indicating that it is a normal insulator (NI). While at 2% strain, the band inversion sustains with an increased global bandgap (31 meV), and the calculation of the $Z2$ invariant yielded a value of 1, indicating its nontrivial topological nature. Meanwhile, the spin texture around VBM exhibits notable alteration compared to the monolayer in an unstrained state. Concretely, the spin texture of VBM undergoes a transition, from having the same spin polarization direction on the same side to alternating spin-up and spin-down configurations, similar to that observed in monolayer $CaAlGaTe_4$ and other TIs. The orbital schematic for the inverted bands of $MgGaInTe_4$ monolayers indicates that the variations in the $s$, $p_z$, and $p_{x,y}$ orbitals near the Fermi level, when considering SOC, are the triggers behind the band inversion and hybrid textures. Specifically, the bands contributed by Ga-*s* and Te-$p_z$ orbitals exhibit typical Rashba spin splitting textures, while those contributed by Te-$p_{x,y}$ orbitals display spin alternating anomalies and *out-of-plane* spin components. In Janus TIs, the *out-of-plane* spin polarization ($S_z \neq 0$) arises from higher-order terms in the $k \cdot p$ Hamiltonian, a direct consequence of inversion symmetry breaking and strong SOC. While the conventional linear Rashba model ($H_R \propto \alpha_R(\sigma_x k_y - \sigma_y k_x)$) confines spin polarization to the *in-plane* directions ($S_x$, $S_y$), symmetry lowering in Janus structures enables nonlinear terms such as $\lambda(k_x^3 - 3k_x k_y^2)\sigma_z$ [17,18]. These terms break the strict linear spin-momentum locking and allow $S_z$ to emerge. In MAA'$Z_xZ'_{(4-x)}$ systems, this effect is amplified by *s*-*p* orbital hybridization near the Fermi level, which enhances momentum-dependent SOC. The interplay between atomic asymmetry (such as the substitution of Te and S) and strain-induced lattice distortion affects the appearance of $S_z$, as demonstrated in the hybrid spin textures of $MgGaInTe_4$ under biaxial strain (Figure S7). It is noteworthy that similar strain-induced changes also occur during the doping process of the MAA'$Z_xZ'_{(4-x)}$ systems, exemplified by the monolayers of $MgAlGaTe_4$ and $CaAlGaTe_4$, as shown in Figure S11.

For these Janus MAA'$Z_xZ'_{(4-x)}$ TIs, in addition to the Rashba effect at VBM caused by band inversion in monolayers such as $MgGaInTe_4$, ideal Rashba splitting also occurs at CBM in $SrGa_2Te_3Se$ monolayers, as shown in Figure S10. By examining the band structures of monolayers composed of similar elements, we discover that the Rashba splitting arises from the continuous shift of the conduction band towards the Fermi level during heavy doping. This movement causes the originally deep-level Rashba splitting to eventually relocate to the CBM, transforming into an ideal Rashba splitting.

Finally, the Schematic diagram of the electronic structure characteristics of all monolayers in the MAA'$Z_xZ'_{(4-x)}$ systems indicates that the emergence of the Rashba effect and topological states is closely related to the elemental composition. For $Z_xZ'_{(4-x)}$ being $S_3Se$ and $S_3Te$, the MAA'$Z_xZ'_{(4-x)}$ monolayers are all semiconductor materials, with approximately 62% being Rashba semiconductors; for $Z_xZ'_{(4-x)}$ being $Se_3S$ and $Se_3Te$, 70% of the monolayers are Rashba semiconductors, and 25% of the monolayers (such as $CaGa_2Se_3S$, etc.) begin to exhibit topological properties, with the $CaGa_2Te_3S$ monolayer also showing metallic behavior; for $Z_xZ'_{(4-x)}$ being $Te_3S$ and $Te_3Se$, 75% of the monolayers are TIs, with the remainder being metals; for $Z_xZ'_{(4-x)}$ being $S_2Se_2$, half of the monolayers are Rashba semiconductors, with one TI ($SrGa_2S_2Se_2$) and two metals ($CaGa_2S_2Se_2$, $SrGaAlS_2Se_2$); for $Z_xZ'_{(4-x)}$ being $S_2Te_2$ and $Se_2Te_2$, 90% of the monolayers are TIs (except for $SrGa_2S_2Te_2$ and $SrGaAlS_2Te_2$, which are metals); for $Z_xZ'_{(4-x)}$ being $S_4$, all monolayers are conventional semiconductors, for $Se_4$, 77% are Rashba semiconductors (except for $CaGa_2Se_4$ and $SrGa_2Se_4$, which are metals and TIs, respectively), for $Te_4$, 55% are TIs, with the rest being metals or Rashba semiconductors.

**Table S7**. Bader charge transfer in $MgAA'Z_xZ'_{(4-x)}$ monolayers with Rashba effect. Subscripts 'bot' and 'top' denote the bottom and top atomic layers, respectively.

| | $q1_{bot}$ (e) | $q2$ (e) | $q3$ (e) | $q4$ (e) | $q5$ (e) | $q6$ (e) | $q7_{top}$ (e) | $Q_{q1\text{-}q7}$ (e) | $\alpha_R$ eVÅ |
|---|---|---|---|---|---|---|---|---|---|
| $MgAl_2S_4$ | -1.484 | 2.191 | -1.514 | 1.613 | -1.513 | 2.191 | -1.484 | 0 | / |
| $MgAlGaSe_4$ | -0.654 | 0.945 | -1.086 | 1.582 | -1.447 | 2.040 | -1.380 | 0.726 | 0.223 |
| $MgAlGaTe_4$ | -0.425 | 0.603 | -0.951 | 1.536 | -1.339 | 1.779 | -1.202 | 0.777 | 0.894 |
| $MgAlInTe_4$ | -0.497 | 0.698 | -0.973 | 1.533 | -1.322 | 1.753 | -1.182 | 0.685 | 0.131 |
| $MgGaInSe_4$ | -0.680 | 0.979 | -1.092 | 1.583 | -1.074 | 0.937 | -0.653 | 0.027 | 0.164 |
| $MgAlGaS_2Se_2$ | -0.653 | 0.957 | -1.084 | 1.598 | -1.519 | 2.175 | -1.472 | 0.819 | 0.153 |
| $MgAlGaS_3Se$ | -0.658 | 1.019 | -1.170 | 1.613 | -1.509 | 2.183 | -1.478 | 0.82 | 0.326 |
| $MgAlGaSe_3S$ | -0.652 | 0.946 | -1.089 | 1.585 | -1.453 | 2.132 | -1.469 | 0.817 | 0.278 |
| $MgGaAlSe_3S$ | -0.806 | 1.104 | -1.094 | 1.582 | -1.451 | 2.051 | -1.386 | 0.58 | 0.035 |
| $MgGaAlSe_3Te$ | -0.652 | 0.937 | -1.083 | 1.587 | -1.447 | 1.916 | -1.257 | 0.605 | 0.181 |
| $MgAl_2Se_3Te$ | -1.259 | 1.919 | -1.453 | 1.588 | -1.448 | 2.017 | -1.363 | 0.104 | 0.03 |
| $MgGa_2Se_3S$ | -0.652 | 0.945 | -1.082 | 1.580 | -1.088 | 1.103 | -0.907 | 0.255 | 0.159 |
| $MgGa_2Se_3Te$ | -0.425 | 0.707 | -0.652 | 1.584 | -1.077 | 0.936 | -0.652 | 0.227 | 0.073 |

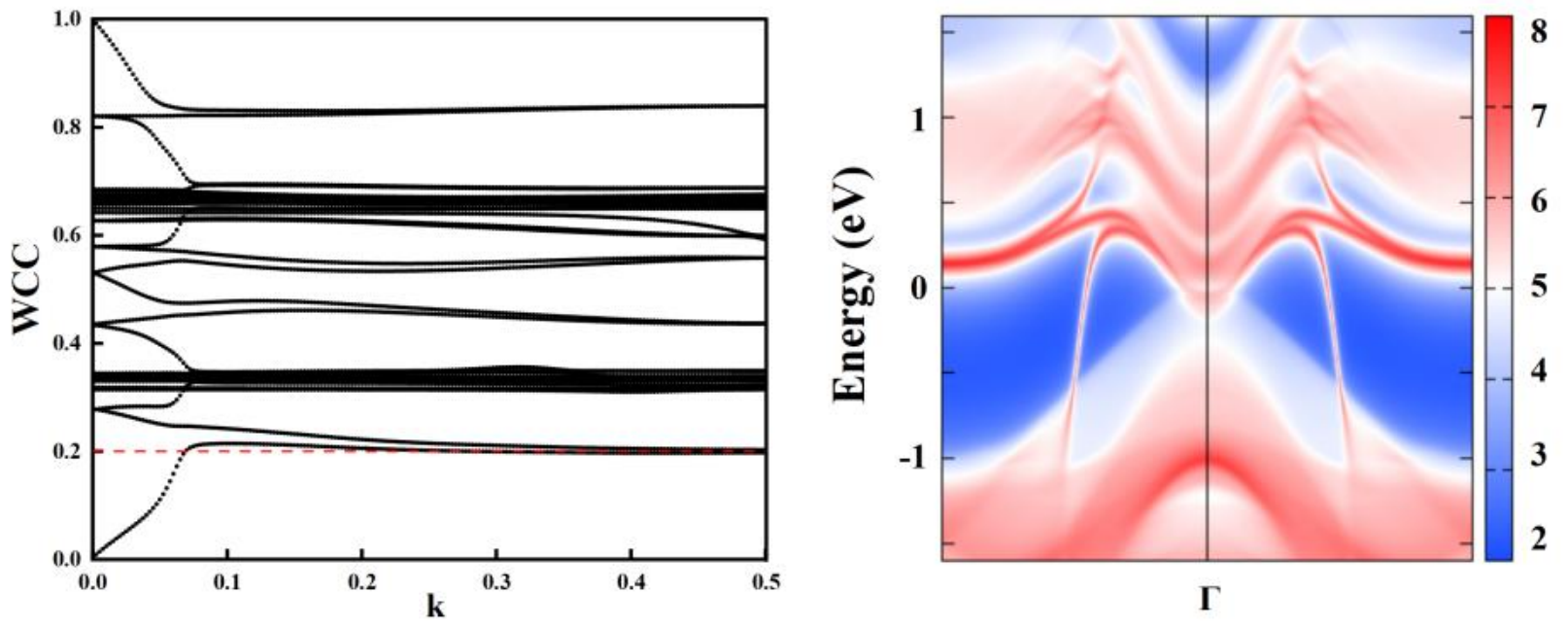

**Figure S6**. The WCC evolution and edge state of the $CaAlGaTe_4$ monolayer.

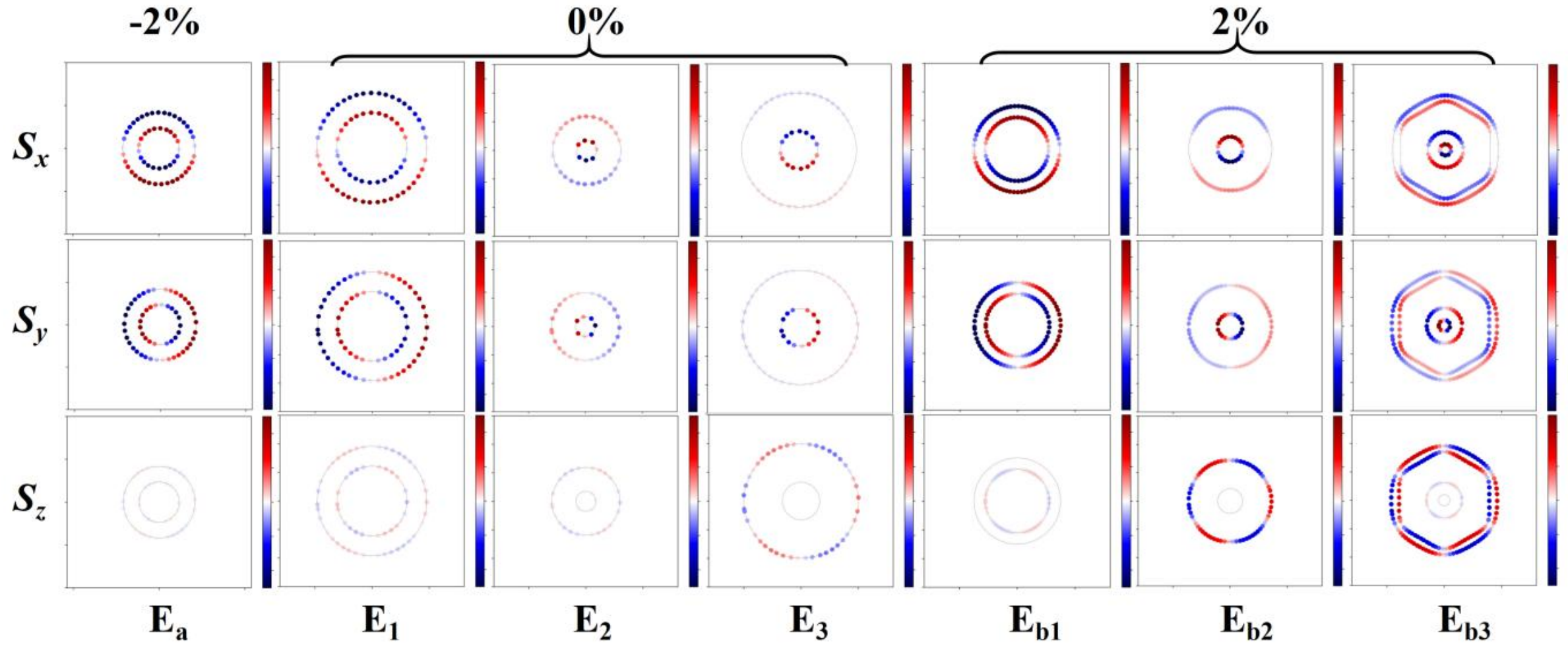

**Figure S7**. Spin texture evolution of the $MgGaInTe_4$ monolayer at selected energy levels under biaxial strains (−2% to 2%).

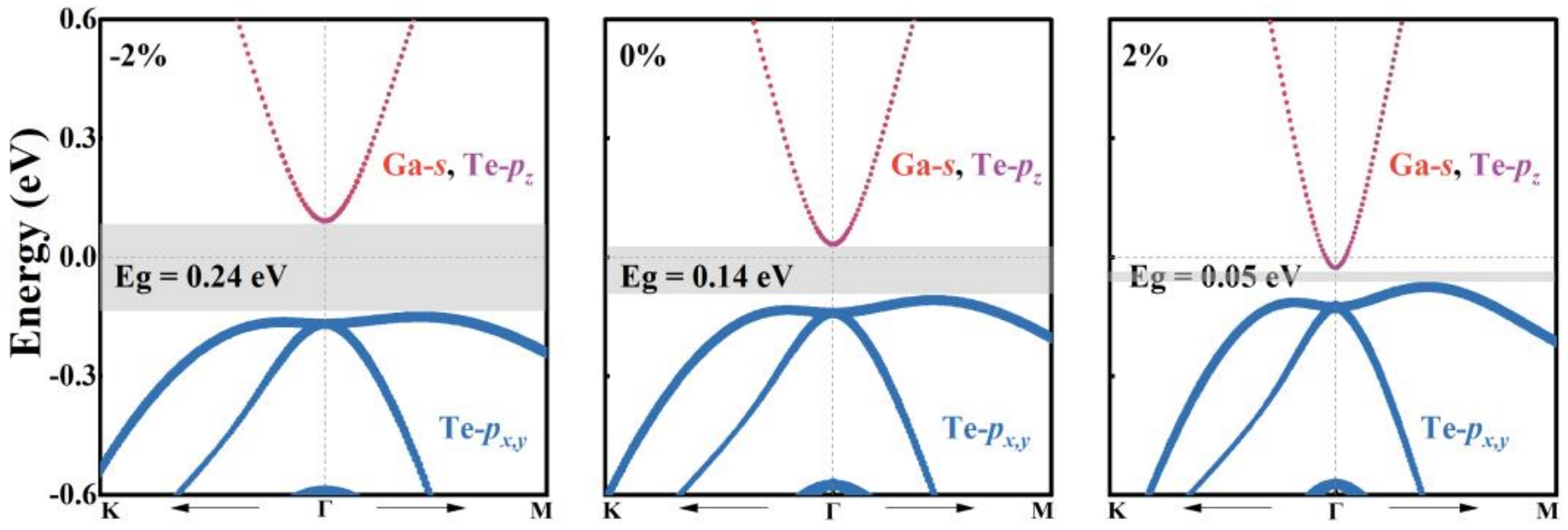

**Figure S8**. Projected band structures of $MgGaInTe_4$ monolayer under -2% to 2% biaxial strain when SOC is not considered.

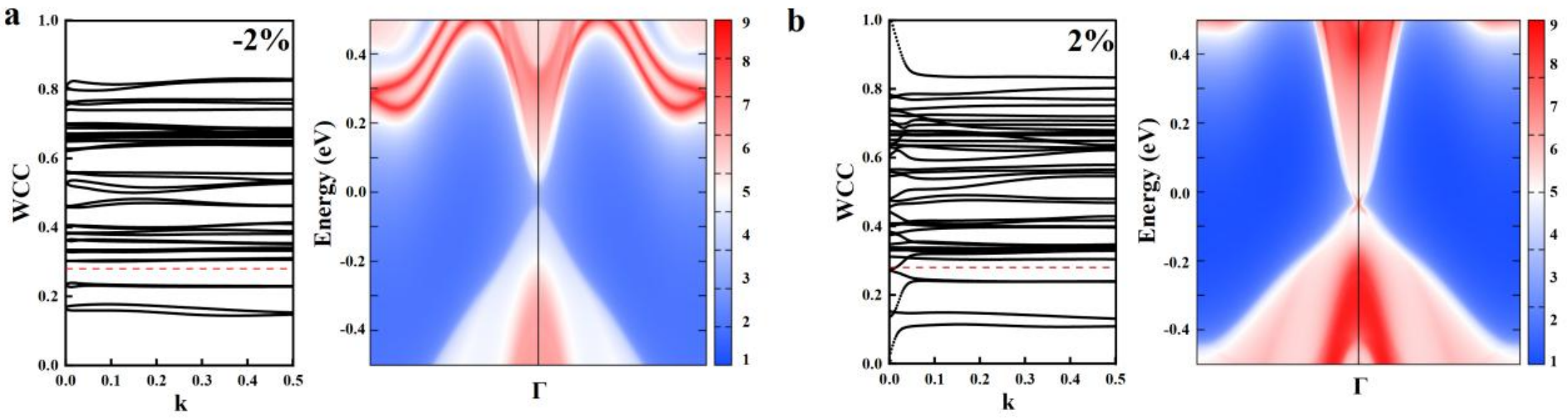

**Figure S9**. The WCC evolution and edge states of $MgGaInTe_4$ monolayer under biaxial strains of -2% and 2%.

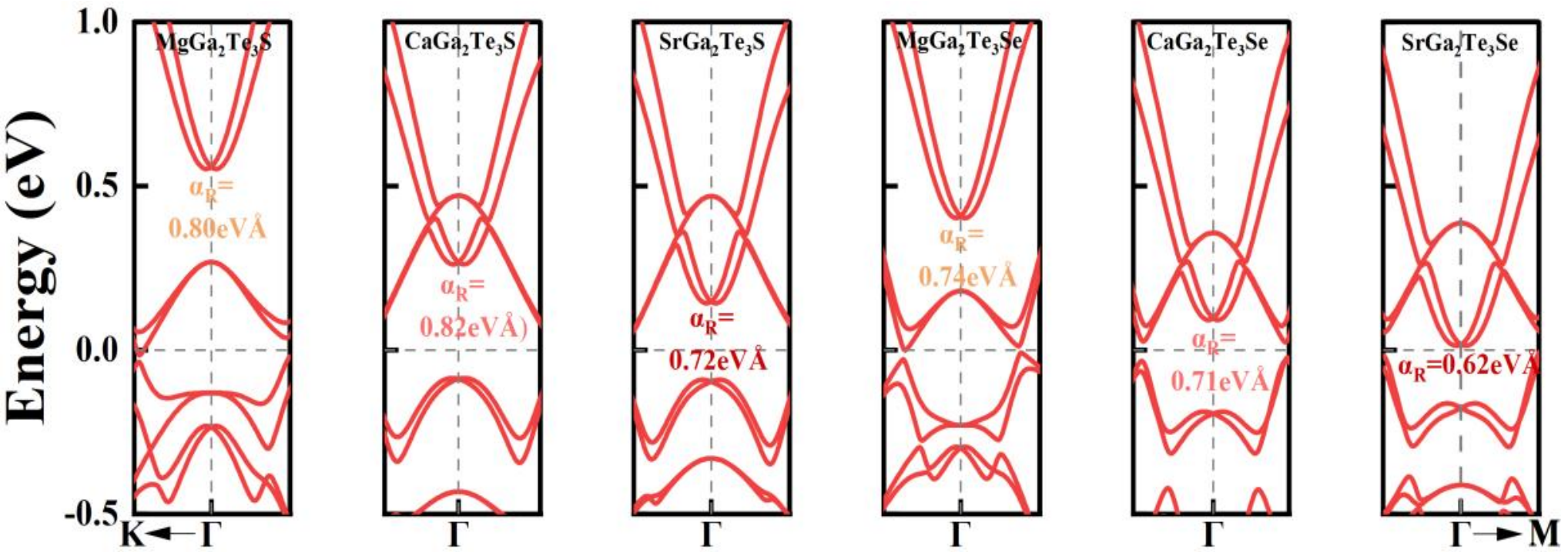

**Figure S10**. Band structures of $MGa_2Te_3S$ monolayers and $MGa_2Te_3Se$ monolayers when SOC is considered.

| | | | | | | | | | | | | | |
|---|---|---|---|---|---|---|---|---|---|---|---|---|---|
| Al-$s$ /S-$p_{xy}$ | Al-$s$ /Te-$p_{xy}$ | Al-$s$ /Se-$p_{xy}$ | Al-$s$ ,Se-$p_z$/Se-$p_{xy}$ | Te-$p_{xy}$ | Te-$p_{xy}$/ Al-s | **$MgAl_2$** | Al-$s$ /Se-$p_{xy}$ | Te-$p_{xy}$ | Te-$p_{xy}$/Al-$s$ | **MgAlGa** | Ga-$s$ /S-$p_{xy}$ | Ga-$s$ ,Se-$p_z$/Se-$p_{xy}$ | Ga-$s$ ,Te-$p_z$/Te-$p_{xy}$ |
| Al-$s$ /S-$p_{xy}$ | Al-$s$ /Te-$p_{xy}$ | Al-$s$ /Se-$p_{xy}$ | Al-$s$ ,Se-$p_z$/Se-$p_{xy}$ | Te-$p_{xy}$ | Te-$p_{xy}$ | **$CaAl_2$** | Al-$s$ /Se-$p_{xy}$ | Te-$p_{xy}$ | Te-$p_{xy}$/Al-$s$ | **CaAlGa** | Ga-$s$ /S-$p_{xy}$ | Ga-$s$ ,Se-$p_z$/Se-$p_{xy}$ | Te-$p_{xy}$/Ga-s |
| Al-$s$ /S-$p_{xy}$ | Al-$s$ /Te-$p_{xy}$ | Al-$s$ /Se-$p_{xy}$ | Al-$s$ ,Se-$p_z$/Se-$p_{xy}$ | Te-$p_{xy}$ | Te-$p_{xy}$ | **$SrAl_2$** | Al-$s$ /Se-$p_{xy}$ | Te-$p_{xy}$ | Te-$p_{xy}$/Al-$s$ | **SrAlGa** | Ga-$s$ /S-$p_{xy}$ | Ga-$s$ ,Se-$p_z$/Se-$p_{xy}$ | Te-$p_{xy}$/Ga-s |
| Ga-$s$ /S-$p_{xy}$ | Ga-$s$ /Te-$p_{xy}$ | Ga-$s$ ,Se-$p_z$/Se-$p_{xy}$ | Ga-$s$ ,Se-$p_z$/Se-$p_{xy}$ | \ | \ | **$MgGa_2$** | Ga-$s$ /Se-$p_{xy}$ | Te-$p_{xy}$ | Te-$p_{xy}$ | **MgAlIn** | In-$s$ ,S-$p_z$/S-$p_{xy}$ | In-$s$ ,Se-$p_z$/Se-$p_{xy}$ | Al-$s$ ,Te-$p_z$/Te-$p_{xy}$ |
| Ga-$s$ /S-$p_{xy}$ | Ga-$s$ /Te-$p_{xy}$ | Se-$p_{xy}$/Ga-$s$ | \ | Ga-$s$ /Te-$p_{xy}$ | Ga-$s$ /Te-$p_{xy}$ | **$CaGa_2$** | \ | Te-$p_{xy}$ | Te-$p_{xy}$ | **CaAlIn** | In-$s$ /S-$p_{xy}$ | In-$s$ ,Se-$p_z$/Se-$p_{xy}$ | \ |
| Ga-$s$ /S-$p_{xy}$ | Ga-$s$ /Te-$p_{xy}$ | Se-$p_{xy}$/Ga-$s$ | Se-$p_{xy}$/Ga-$s$ | Ga-$s$ /Te-$p_{xy}$ | Ga-s,Te-$p_z$/Te-$p_{xy}$ | **$SrGa_2$** | Se-$p_{xy}$/Ga-$s$ | \ | Te-$p_{xy}$ | **SrAlIn** | In-$s$ /S-$p_{xy}$ | In-$s$ ,Se-$p_z$/Se-$p_{xy}$ | Te-$p_{xy}$/Al- |
| **$S_3Se$** | **$S_3Te$** | **$Se_3S$** | **$Se_3Te$** | **$Te_3S$** | **$Te_3Se$** | **$MA_2$ $Z_3Z'$ $Z_2Z'_2$ MAA'** | **$S_2Se_2$** | **$S_2Te_2$** | **$Se_2Te_2$** | **MgGaIn** | Ga-$s$ /S-$p_{xy}$ | Ga-$s$ ,Se-$p_z$/Se-$p_{xy}$ | Te-$p_{x,y}$/Ga-$s$ ,Te-$p_z$ |
| Ga-$s$ ,Se-$p_z$/S-$p_{xy}$ | Al-$s$ /Te-$p_{xy}$ | Ga-$s$ ,Se-$p_z$/Se-$p_{xy}$ | Al-$s$ /Se-$p_{xy}$ | Te-$p_{xy}$ | Te-$p_{xy}$/Al-$s$ | **MgAlGa** | Ga-$s$ ,Se-$p_z$/Se-$p_{xy}$ | Te-$p_{xy}$ | Te-$p_{xy}$/Al-$s$ | **CaGaIn** | Ga-$s$ /S-$p_{xy}$ | \ | Te-$p_{x,y}$/Ga-$s$ ,Te-$p_z$ |
| Ga-$s$ ,Se-$p_z$/S-$p_{xy}$ | Al-$s$ /Te-$p_{xy}$ | Ga-$s$ ,Se-$p_z$/Se-$p_{xy}$ | Al-$s$ /Se-$p_{xy}$ | Ga-$s$ /Te-$p_{xy}$ | \ | **CaAlGa** | Ga-$s$ ,Se-$p_z$/Se-$p_{xy}$ | Te-$p_{xy}$ | Te-$p_{xy}$/Al-$s$ | **SrGaIn** | Ga-$s$ /S-$p_{xy}$ | Se-$p_{xy}$/Ga-s | In-$s$ ,Te-$p_z$/Te-$p_{xy}$ |
| Ga-$s$ ,Se-$p_z$/S-$p_{xy}$ | Al-$s$ /Te-$p_{xy}$ | Ga-$s$ ,Se-$p_z$/Se-$p_{xy}$ | Al-$s$ /Se-$p_{xy}$ | Ga-$s$ /Te-$p_{xy}$ | \ | **SrAlGa** | Ga-$s$ ,Se-$p_z$/Se-$p_{xy}$ | Te-$p_{xy}$ | Te-$p_{xy}$ | **MAA'$Z_4$** | **$S_4$** | **$Se_4$** | **$Te_4$** |
| Ga-s /S-$p_{xy}$ | Ga-$s$ /Te-$p_{xy}$ | Ga-$s$ ,Se-$p_z$/Se-$p_{xy}$ | Ga-$s$ ,Se-$p_z$/Se-$p_{xy}$ | \ | \ | **MgGaAl** | Ga-$s$ /Se-$p_{xy}$ | Te-$p_{xy}$ | Te-$p_{xy}$ | | | | |
| Ga-s /S-$p_{xy}$ | Ga-$s$ /Te-$p_{xy}$ | Se-$p_{xy}$/Ga-s | Ga-$s$ ,Se-$p_z$/Se-$p_{xy}$ | Te-$p_{xy}$ | Te-$p_{xy}$ | **CaGaAl** | Ga-$s$ /Se-$p_{xy}$ | Te-$p_{xy}$ | Te-$p_{xy}$ | | | | |
| Ga-s /S-$p_{xy}$ | Ga-$s$ /Te-$p_{xy}$ | Se-$p_{xy}$ | Se-$p_{xy}$/Ga-s | Te-$p_{xy}$ | Te-$p_{xy}$ | **SrGaAl** | \ | \ | Te-$p_{xy}$ | | | | |

General semiconductor Small Rashba split

Ideal Rashba split TIs

TIs with Rashba splitting at VBM

TIs with Rashba splitting at CBM

**Figure S11**. The orbital contributions at the CBM and VBM at the Γ point in monolayer Janus MAA'$Z_xZ'_{(4-x)}$ family. The sections before and after the slash in the box represent the orbital contributions of the CBM and VBM, respectively, or indicate that both have the same orbital contribution.